\documentclass[a4paper,11pt]{article}
\pdfoutput=1 

\usepackage{jcappub} 

\usepackage[T1]{fontenc} 
\usepackage{graphicx}
\graphicspath{{figures/}}
\usepackage[utf8]{inputenc}
\usepackage{amsmath}
\usepackage{amsfonts}
\usepackage{amssymb}
\usepackage{enumitem}
\usepackage{multirow}
\usepackage{orcidlink}
\usepackage{CJKutf8}
\usepackage{color}
\usepackage[capitalise]{cleveref}
\usepackage{acro}
\usepackage{adjustbox}
\usepackage{array}
\usepackage{natbib}
\usepackage{dblfnote}
\DFNalwaysdouble
\usepackage{slashed}
\usepackage{dcolumn}
\usepackage{hyperref}
\usepackage{geometry}
\usepackage{subfig} 
\usepackage{overpic}
\usepackage{inputenc}
\usepackage{caption}
\usepackage[theorems,skins]{tcolorbox}

\def\({\left(}
\def\){\right)}
\def\[{\left[}
\def\]{\right]}
\def\be{\begin{eqnarray}}
\def\ee{\end{eqnarray}}

\DeclareAcronym{GW}{
  short = GW ,
  long = gravitational wave ,
  short-plural = s 
}
\DeclareAcronym{LIGO}{
  short = LIGO ,
  long = Laser Interferometer Gravitational-wave Observatory ,
  short-plural = 
}

\DeclareAcronym{BBN}{
  short = BBN ,
  long = Big Bang Nucleosynthesis  ,
  short-plural = 
}

\DeclareAcronym{LISA}{
  short = LISA ,
  long = Laser Interferometer Space Antenna ,
  short-plural =  
}
\DeclareAcronym{SKA}{
  short = SKA ,
  long = Square Kilometre Array ,
  short-plural =  
}  
  
\DeclareAcronym{SNR}{
	short = SNR ,
	long = signal-to-noise ratio ,
	short-plural = 
}

\DeclareAcronym{PTA}{
	short = PTA ,
	long = pulsar timing array ,
	short-plural = 
}

\DeclareAcronym{FLRW}{
  short = FLRW ,
  long = Friedmann-Lemaitre-Robertson-Walker ,
  short-plural =  
}

\DeclareAcronym{SIGW}{
	short = SIGW ,
	long = scalar induced gravitational wave ,
	short-plural =  s
}

\DeclareAcronym{PBH}{
	short = PBH ,
	long = primordial black hole ,
	short-plural =  s
}

\DeclareAcronym{CMB}{
	short = CMB ,
	long = cosmic microwave background ,
	short-plural =  
}
\DeclareAcronym{DM}{
	short = DM ,
	long = dark matter ,
	short-plural =  
}

\DeclareAcronym{SGWB}{
	short = SGWB ,
	long = stochastic gravitational	wave background ,
	short-plural =  s
}

\DeclareAcronym{LSS}{
	short = LSS ,
	long = large-scale structure ,
	short-plural =  
}

\DeclareAcronym{EFA}{
	short = EFA ,
	long = effective fluid approach,
	short-plural =  
}

\DeclareAcronym{RD}{
	short = RD ,
	long = radiation-dominated ,
	short-plural =  
}

\DeclareAcronym{SMBHB}{
  short = SMBHB ,
  long = supermassive black hole binary ,
}

\title{\boldmath Scalar induced gravitational waves in $f(R)$ gravity}

\author[a,1]{Jing-Zhi Zhou,\note{Corresponding author.} \orcidlink{0000-0003-2792-3182} }
\author[b,c]{Yu-Ting Kuang, \orcidlink{0000-0002-7431-4454}}
\author[d,b]{Di Wu,\orcidlink{0000-0001-7309-574X}}
\author[a]{Fei-Yu Chen,\orcidlink{0000-0002-8674-9316}}
\author[a,e]{ H. L\"u,\orcidlink{0000-0001-7100-2466}}
\author[b,c]{Zhe Chang \orcidlink{0000-0002-9720-803X}}

\affiliation[a]{Center for Joint Quantum Studies and Department of Physics,
School of Science, Tianjin University, Tianjin 300350, China}
\affiliation[b]{Institute of High Energy Physics, Chinese Academy of Sciences, Beijing 100049, China}
\affiliation[c]{University of Chinese Academy of Sciences, Beijing 100049, China}
\affiliation[d]{School of Fundamental Physics and Mathematical Sciences, Hangzhou Institute for Advanced Study, University of Chinese Academy of Sciences, Hangzhou 310024, China}
\affiliation[e]{Joint School of National University of Singapore and Tianjin University, International Campus of Tianjin University, Binhai New City, Fuzhou 350207, China}

\emailAdd{zhoujingzhi@ihep.ac.cn}

\abstract{ We investigate the first and second order cosmological perturbation equations in $f(R)$ modified gravity theory and provide the equation of motion of second order scalar induced gravitational waves. We find that the effects of modified gravity not only change the form of the equation of motion of second order scalar induced gravitational waves but also contribute an additional anisotropic stress tensor, composed of first order scalar perturbations, to the source term of the gravitational waves. We calculate the energy density spectrum of second order scalar induced gravitational waves in the HS model. Utilizing current pulsar timing array observational data, we perform a rigorous Bayesian analysis of the parameter space of the HS model.

}

\begin{document}

\maketitle
\flushbottom

\section{Introduction}\label{sec:1}
The detection of \acp{GW} from the mergers of black holes or neutron stars by LIGO/VIRGO has inaugurated the era of gravitational wave astronomy \cite{LIGOScientific:2016aoc,LIGOScientific:2016vlm}. Over the past decade, gravitational waves have become a major research topic in cosmology, astrophysics, and astronomy \cite{Annala:2017llu,LIGOScientific:2017adf,BICEP:2021xfz}. 

In this paper, we focus on gravitational waves induced by primordial scalar perturbations in cosmology, which can serve as probes of the inflationary period and the thermal history of cosmic evolution \cite{Ananda:2006af,Domenech:2021ztg,Kohri:2018awv,Mollerach:2003nq}. The information from quantum fluctuations during inflation and the subsequent physical processes of cosmic evolution are imprinted in the induced gravitational waves observed today. In June 2023,  several international \ac{PTA} collaborations, such as NANOGrav \cite{NANOGrav:2023gor}, PPTA \cite{Reardon:2023gzh}, EPTA \cite{EPTA:2023fyk}, and CPTA \cite{Xu:2023wog}, provided evidence for the existence of a \ac{SGWB} in the nHz frequency range. Scalar induced gravitational waves (SIGWs), as one of the most plausible sources of this background, have received extensive attention.  The researches on  SIGWs have been extended to \acp{PBH} \cite{Wang:2019kaf,Byrnes:2018txb,Inomata:2020lmk,Ballesteros:2020qam,Lin:2020goi,Chen:2019xse,Cai:2019elf,Cai:2019jah,Ando:2018qdb,Di:2017ndc,Gao:2021vxb,Changa:2022trj,Zhou:2020kkf,Cai:2021wzd}, gauge issue \cite{Hwang:2017oxa,Yuan:2019fwv,Inomata:2019yww,DeLuca:2019ufz,Domenech:2020xin,Chang:2020tji,Ali:2020sfw,Lu:2020diy,Tomikawa:2019tvi,Gurian:2021rfv,Uggla:2018fiy,Ali:2023moi}, different epochs of the Universe \cite{Papanikolaou:2020qtd,Domenech:2020kqm,Domenech:2019quo,Inomata:2019zqy,Inomata:2019ivs,Witkowski:2021raz,Dalianis:2020gup,Hajkarim:2019nbx,Bernal:2019lpc,Das:2021wad,Haque:2021dha,Domenech:2020ssp,Domenech:2021and,Liu:2023pau}, damping effect  \cite{Mangilli:2008bw,Saga:2014jca,Zhang:2022dgx,Yuan:2023ofl,Yu:2024xmz}, primordial non-Gaussianity  \cite{Cai:2018dig,Atal:2021jyo,Zhang:2020uek,Yuan:2020iwf,Davies:2021loj,Rezazadeh:2021clf,Kristiano:2021urj,Bartolo:2018qqn,Adshead:2021hnm,Li:2023qua,Li:2023xtl,Garcia-Saenz:2022tzu}, and higher order correction of SIGWs \cite{Zhou:2021vcw,Chang:2023vjk,Wang:2023sij,Chang:2022nzu,Chang:2022dhh,Zhou:2024ncc}.

The experimental observation of the \ac{SGWB} provides a new window for studying SIGWs and exploring potential new physics in the process of cosmic evolution. In cosmology, numerous new physical models are based on modified gravity theories \cite{Koyama:2015vza}. Testing these cosmological models based on modified gravity theories is undoubtedly a crucial issue in the studies of modified gravity and cosmology. Utilizing current PTA observational data, we can test various modified gravity models and constrain their parameter spaces. This paper systematically investigates the impact of modified gravity effects in $f(R)$ gravity theory on second order SIGWs and PTA observations. Specifically, through the effective fluid approach \cite{Arjona:2018jhh}, we thoroughly investigate the equations of motion for first order cosmological perturbations and second order SIGWs in $f(R)$ modified gravity theory. We present the general form of the equations of motion for second order SIGWs in $f(R)$ theory and their formal solutions. These results are independent of the primordial power spectrum and the specific form of the $f(R)$ modified gravity theory. We apply these general results to the HS model and calculate the energy density spectrum of second order SIGWs. Our results reveal that the effects of $f(R)$ theory impact the energy density spectrum of second order SIGWs.  Based on the these results, we can constrain the parameter space of the HS model using current PTA observation data \cite{Hu:2007nk}.

This paper is organized as follows. In Sec.~\ref{sec:2}, we review the basics of second order SIGWs and provide detailed calculations for first and second order cosmological perturbations. In Sec.~\ref{sec:3} we investigate the first and second order cosmological perturbations in $f(R)$ theory. We derive the equation of motions for second order SIGWs in $f(R)$ theory and present the corresponding formal solutions. In Sec.~\ref{sec:4}, we present the general formula for calculating the energy density spectrum of second order SIGWs in $f(R)$ theory and investigate the energy density spectrum in HS model. We perform a Bayesian analysis by combining \ac{PTA} observational data to constrain the parameter space of the HS model. Finally, we summarize our results and give some discussions in Sec.~\ref{sec:5}.

\section{Scalar induced gravitational waves in general relativity}\label{sec:2}
In this section, we review the main results of second order SIGWs and analyze the impact of the anisotropic stress tensor $\Pi_{ij}$ on the cosmological perturbations. The perturbed metric in the flat \ac{FLRW} spacetime with Newtonian gauge is given by
\begin{equation}\label{eq:ds}
\begin{aligned}
	\mathrm{d} s^2  =a^2(\eta)\left[-\left(1+2 \phi^{(1)}\right) \mathrm{d} \eta^2 +\left(\left(1-2 \psi^{(1)}\right) \delta_{i j}+\frac{1}{2} h_{i j}^{(2)}\right) \mathrm{d} x^i \mathrm{~d} x^j\right] \ ,
\end{aligned}
\end{equation}
where $\eta$ is the conformal time. $\phi^{(1)}$ and $\psi^{(1)}$ represent first order scalar perturbations. $h^{(2)}_{ij}$ represents second order tensor perturbation. We have neglected the first order vector and tensor perturbations. To derive the equation of motion of second order SIGWs, we need to solve the perturbations of the Einstein field equation in the \ac{FLRW} spacetime order by order. Specifically, by considering the energy-momentum tensor of a perfect fluid, the Einstein field equation can be formulated as follows:
\begin{equation}\label{eq:Ee}
      R_{\mu \nu}-\frac{1}{2} g_{\mu \nu} R=8 \pi G T^M_{\mu\nu} \ ,
\end{equation}
where $T^M_{\mu\nu}=\left(\rho +p \right)u_\mu u_\nu + p g_{\mu\nu}$ is the energy-momentum tensor of the perfect fluid. By substituting Eq.~(\ref{eq:ds}) into Eq.~(\ref{eq:Ee}), we can obtain the equations of motion of $a(\eta)$, $\phi^{(1)}$, $\psi^{(1)}$, and $h_{ij}^{(2)}$. In the rest of this section, we will investigate the cosmological perturbations in the \ac{FLRW} spacetime order by order.

\subsection{Background evolution}\label{sec:2.1}
The flat \ac{FLRW} metric is given by
\begin{equation}\label{eq:ds0}
\begin{aligned}
	\mathrm{d} s^2  =a^2(\eta)\left[- \mathrm{d} \eta^2 + \delta_{i j} \mathrm{d} x^i \mathrm{d} x^j\right] \ .
\end{aligned}
\end{equation}
The $0$-th order perturbation of the energy-momentum tensor of a perfect fluid can be represented as
\begin{eqnarray}\label{eq:T0}
    T^{M,(0)}_{\mu\nu}=\left(1 +w \right)\rho^{(0)} u^{(0)}_\mu u^{(0)}_\nu + w \rho^{(0)} g_{\mu\nu}^{(0)} \ ,
\end{eqnarray}
where $w=p^{(0)}/\rho^{(0)}$ is the equation of state. Here, $u^{(0)}_\mu=\left( -a(\eta),0,0,0 \right)$. By substituting Eq.~(\ref{eq:ds0}) and Eq.~(\ref{eq:T0}) into Eq.~(\ref{eq:Ee}), we obtain
\begin{eqnarray}\label{eq:G0}
G^{(0)}_{00}&=&\kappa T^{M,(0)}_{00}:~~ 3 \mathcal{H}^2=\kappa a^2 \rho^{(0)}  \ , \nonumber\\ 
G^{(0)}_{ij}&=&\kappa T^{M,(0)}_{ij}:~~ \delta_{ij} \left( \mathcal{H}^2+2\mathcal{H}' \right)=-\kappa \delta_{ij}  a^2 w \rho^{(0)} \ ,
\end{eqnarray}
where $\kappa=8\pi G$. The prime denotes the derivative with respect to
the conformal time $\eta$. Eq.~(\ref{eq:G0}) allows us to express the derivative of the comoving Hubble parameter and the $0$-th order density perturbation as 
\begin{eqnarray}\label{eq:r0}
    \mathcal{H}' =-\frac{1}{2}(1+3 w) \mathcal{H}^2 \ , \  \rho^{(0)}=3 \mathcal{H}^2/\kappa a^2 \ .
\end{eqnarray}

\subsection{First order scalar perturbations}\label{sec:2.2}
As we mentioned, we have ignored the first order vector and tensor perturbation in Eq.~(\ref{eq:ds}). Since the vector perturbation decay as $1/a^2$, it is generally difficult to produce large-amplitude primordial vector perturbations during the inflationary period, unless a special inflationary model is constructed \cite{Graham:2015rva,Okano:2020uyr}. For first order tensor perturbations, there are no strong observational constraints on small scales, and there are many methods to construct large-amplitude primordial tensor perturbations on small scales \cite{Cai:2020ovp,Gorji:2023ziy}. In this case, large-amplitude primordial tensor perturbations will have a significant impact on the induced gravitational wave background in the high-frequency region \cite{Chang:2022vlv}. Here, we ignore the effects of primordial vector and tensor perturbations, and only consider the large-amplitude primordial curvature perturbations on small scales and their effects on cosmological perturbation evolution. By substituting Eq.~(\ref{eq:ds}) into Eq.~(\ref{eq:Ee}) and ignoring the second order perturbations, we obtain the first order perturbations of the Einstein field equation in \ac{FLRW} spacetime 
\begin{eqnarray}
G^{(1)}_{00}&=&\kappa T^{M,(1)}_{00}:~~ -6\mathcal{H}\psi^{(1)'}+2\Delta\psi^{(1)}=\kappa a^2\left( \rho^{(1)}+2\rho^{(0)}\phi^{(1)}\right)  \ , \label{eq:G100}\\ 
G^{(1)}_{0i}&=&\kappa T^{M,(1)}_{0i}:~~ 2\left( \mathcal{H} \partial_i \phi^{(1)}+\partial_i \psi^{(1)'} \right)=-\kappa\left(1+w \right)a^2 \rho^{(0)} u^{(1)}_i \ , \label{eq:G10i}\\  
G^{(1)}_{ij}&=&\kappa T^{M,(1)}_{ij}:~~ \delta_{ij} \left( 2\mathcal{H}\left( \phi^{(1)}+\psi^{(1)}\right)+4\mathcal{H}'\left(\phi^{(1)}+\psi^{(1)} \right)+2\mathcal{H}\left( \phi^{(1)'}+2\psi^{(1)'} \right)+2\psi^{(1)''} \right. \nonumber\\
& &\left.+\Delta\phi^{(1)}-\Delta\psi^{(1)} \right)+\partial_i\partial_j\psi^{(1)}-\partial_i\partial_j\phi^{(1)}=\kappa a^2 \delta_{ij} \left( c_s^2\rho^{(1)}-2w\rho^{(0)}\psi^{(1)} \right) \ , \label{eq:G1ij}
\end{eqnarray}
where $c_s=\sqrt{p^{(1)}/\rho^{(1)}}$ is the speed of sound. The explicit expressions of the perturbation of energy-momentum tensor can be found in  Ref.~\cite{Inomata:2020cck}. By using Eq.~(\ref{eq:G100}) and Eq.~(\ref{eq:G10i}), we can express first order energy density perturbation $\rho^{(1)}$ and first order velocity perturbation $u^{(1)}_{i}$ as functions of first order scalar perturbations
\begin{eqnarray}
    \rho^{(1)}&=&\frac{1}{\kappa a^2}\left(-6 \mathcal{H}\left(\mathcal{H} \phi^{(1)}+\psi^{(1)^{\prime}}\right)+2 \Delta \psi^{(1)}\right) \ , \label{eq:r1}\\
    u_i^{(1)}&=&-\frac{2}{3(1+w) \mathcal{H}^2}\left(\mathcal{H} \partial_i \phi^{(1)}+\partial_i \psi^{(1)^{\prime}}\right) \ , \label{eq:u1}
\end{eqnarray}
where the relationship in Eq.~(\ref{eq:r0}) is utilized. By substituting Eq.~(\ref{eq:r1}) and Eq.~(\ref{eq:u1}) into Eq.~(\ref{eq:G1ij}), we obtain the equation of motion for first order scalar perturbations without energy density or velocity perturbations
\begin{eqnarray}\label{eq:eom10}
    & &\delta_{ij}\left( 2\mathcal{H} \left(3(c_s^2-w)\mathcal{H}\phi^{(1)}+\phi^{(1)'}+(2+3c_s^2)\psi^{(1)'}   \right)  +2\psi^{(1)''}+\Delta \phi^{(1)}-\Delta \psi^{(1)} \right. \nonumber\\
    & &\left. -2c_s^2\Delta \psi^{(1)}\right)+\partial_i\partial_j\psi^{(1)}-\partial_i\partial_j\phi^{(1)}=0 \ .
\end{eqnarray}
Eq.~(\ref{eq:eom10}) comes from the spatial-spatial component of the first order perturbation equation of the Einstein field equation in \ac{FLRW} spacetime. To solve Eq.~(\ref{eq:eom10}), we need to use the decomposition operator on the FRW spacetime to decompose Eq.~(\ref{eq:eom10}) into two equations. More precisely, an arbitrary three-dimensional spatial tensor field $S_{ij}$ on \ac{FLRW} spacetime can be decomposed into scalar, vector, and tensor modes \cite{Chang:2020tji}
\begin{equation}
	S_{i j}=S_{i j}^{(H)}+2 \delta_{i j} S^{(\Psi)}+2 \partial_{i} \partial_{j} S^{(E)}+\partial_{j} S_{i}^{(C)}+\partial_{i} S_{j}^{(C)} \ .
\end{equation}
We define the following decomposed operators to fulfill this decomposition
\begin{equation}\label{eq:Dec1}
	S_{i j}^{(H)} \equiv\Lambda_{ij}^{kl}S_{k l}=\left(\mathcal{T}_{i}^{k} \mathcal{T}_{j}^{l}-\frac{1}{2} \mathcal{T}_{i j} \mathcal{T}^{k l}\right) S_{k l} \ \ , \ \ S^{(\Psi)} \equiv \frac{1}{4} \mathcal{T}^{k l} S_{k l} \ .
\end{equation}
\begin{equation}\label{eq:Dec2}
	S^{(E)} \equiv \frac{1}{2} \Delta^{-1}\left(\partial^{k} \Delta^{-1} \partial^{l}-\frac{1}{2} \mathcal{T}^{k l}\right) S_{l k} \ \ , \ \ S_{i}^{(C)} \equiv \Delta^{-1} \partial^{l} \mathcal{T}_{i}^{k} S_{l k} \ ,
\end{equation}
where the transverse operator is given by
\begin{equation}
	\mathcal{T}_{j}^{i} \equiv \delta_{i}^{i}-\partial^{i} \Delta^{-1} \partial_{j} \ .
\end{equation} 
It should be emphasized that these decomposition operators are applicable only to \ac{FLRW} spacetime. In arbitrary spacetime, $S_{i}^{(C)}$ and $S^{(E)}$ cannot be distinguished \cite{OMurchadha:1973byk}. 

By utilizing the decomposition operators in Eq.~(\ref{eq:Dec1}) and Eq.~(\ref{eq:Dec2}), Eq.~(\ref{eq:eom10}) can be decomposed into the following two equations
\begin{eqnarray}
  &&\psi^{(1)}-\phi^{(1)}=0 \ , \label{eq:s2}
      \\ \nonumber\\
&&2\mathcal{H} \left(3(c_s^2-w)\mathcal{H}\phi^{(1)}+\phi^{(1)'}+(2+3c_s^2)\psi^{(1)'}   \right)  +2\psi^{(1)''}+\Delta \phi^{(1)}  \label{eq:s1} \nonumber\\
     &&-\Delta \psi^{(1)}-2c_s^2\Delta \psi^{(1)}=0 \ .
\end{eqnarray}
In the case of a perfect fluid, its energy-momentum tensor only contains three unknowns: energy density $\rho^{(n)}$, pressure $p^{(n)}$, and velocity $u_{i}^{(n)}$. By expressing the equation of state and the speed of sound, we can obtain the relationship between pressure (perturbations) and energy density (perturbations): $p^{(0)}=w\rho^{(0)}$ and $p^{(1)}=c_s^2\rho^{(1)}$. By simplifying the time-time and time-space components of the perturbed Einstein field equation, we express energy density perturbations $\rho^{(1)}$ and velocity perturbations $u_i^{(n)}$ as functions of scalar perturbations $\phi^{(1)}$ and $\psi^{(1)}$. As shown in Eq.~(\ref{eq:s1}) and Eq.~(\ref{eq:s2}), for a perfect fluid, all information about the energy-momentum tensor is contained in the parameters $w$ and $c_s^2$. However, for a general (imperfect) fluid, its energy-momentum tensor would produce an anisotropic stress, namely
\begin{eqnarray}
    \Pi_{i j}=\Pi_{i j}^{(1)}+\frac{1}{2} \Pi_{i j}^{(2)}+\cdot\cdot\cdot
\end{eqnarray}
We can use the decomposition operators to decompose the anisotropic stress $\Pi_{i j}$ as
\begin{eqnarray}\label{eq:Pi0}
    \Pi_{i j}^{(n)}=\sigma_{i j}^{\mathrm{TT},(n)}+\frac{1}{2}\left(\partial_i \sigma_j^{(n)}+\partial_j \sigma_i^{(n)}\right)+\left(\partial_i \partial_j-\frac{1}{3} \delta_{i j} \Delta\right) \sigma^{(n)}  \ , 
\end{eqnarray}
where $\partial^i \sigma_i^{(n)}=\partial^i \sigma_{i j}^{\mathrm{TT}(n)}=\delta^{i j} \sigma_{i j}^{\mathrm{TT},(n)}=0$. In this scenario, there will be additional contributions from anisotropic stress tensors $\Pi_{i j}^{(1)}$ on the right-hand side of Eq.~(\ref{eq:G1ij}). At this stage, the parameters $w$ and $c_s^2$ cannot fully capture the effects of the energy-momentum tensor on first order scalar perturbations. The equations of motion of first order scalar perturbations will certainly involve contributions from anisotropic stress tensors. In the study of scalar-induced gravitational waves, there are two sources of anisotropic stress tensors. One is the interaction between  neutrinos and gravitational waves. Specifically, after neutrino decoupling, the freely streaming neutrinos will produce an anisotropic stress tensor \cite{Weinberg:2003ur}. The anisotropic stress tensor produced by neutrinos will lead to a noticeable suppression of the energy density spectrum of SIGWs at frequencies lower than $10^{-10}$Hz \cite{Zhang:2022dgx}. Another source of anisotropic stress tensors originated from modified gravity. In this case, even if we consider the energy-momentum tensor of an perfect fluid, the effects of modified gravity will still contribute a nonzero effective anisotropic stress tensor \cite{Tsujikawa:2007gd,Arjona:2018jhh,Papanikolaou:2021uhe}.

In previous research on SIGWs, the effects of the anisotropic stress tensor were commonly neglected. During the \ac{RD} era ($w=c_s^2=1/3$), Eq.~(\ref{eq:s1}) and Eq.~(\ref{eq:s2}) can be rewritten as
\begin{equation}\label{eq:phi1}
	\begin{aligned}
    \phi^{(1)''}(\mathbf{x},\eta)+ 4\mathcal{H}\ \phi^{(1)'}(\mathbf{x},\eta)-\frac{1}{3}\Delta \phi^{(1)}(\mathbf{x},\eta) &=0 \ .
	\end{aligned}
\end{equation}
By solving Eq.~(\ref{eq:phi1}) in momentum space, we obtain \cite{Kohri:2018awv}
\begin{eqnarray}\label{eq:T1}
    \phi^{(1)}_{\mathbf{k}}(\eta)=\psi^{(1)}_{\mathbf{k}}(\eta)=\frac{2}{3}\zeta_{\mathbf{k}} T_{\phi}(|\mathbf{k}| \eta) \ ,
\end{eqnarray}
where $\zeta_{\mathbf{k}}$ in Eq.~(\ref{eq:T1}) is the primordial curvature perturbation. The transfer function $T_{\phi}(|\mathbf{k}|\eta)$ is given by \cite{Kohri:2018awv}
\begin{eqnarray}\label{eq:T2}
    T_{\phi}(|\mathbf{k}|\eta)=\frac{9}{(|\mathbf{k}|\eta)^{2}}\left(\frac{\sqrt{3}}{|\mathbf{k}|\eta} \sin \left(\frac{|\mathbf{k}| \eta}{\sqrt{3}}\right)-\cos \left(\frac{|\mathbf{k}| \eta}{\sqrt{3}}\right)\right) \ .
\end{eqnarray}

\subsection{Scalar induced gravitational waves}\label{sec:2.3}
After studying first order scalar perturbations, we can calculate the equation of motion of second order gravitational waves. By substituting Eq.~(\ref{eq:ds}) into Eq.~(\ref{eq:Ee}) and extracting the transverse and traceless part of the space-space component of the second order perturbation, we obtain
\begin{eqnarray}\label{eq:eh}
    h_{ij}^{(2)''}\left(\mathbf{x},\eta \right)+2\mathcal{H}h_{ij}^{(2)'}\left(\mathbf{x},\eta \right)-\Delta h_{ij}^{(2)}\left(\mathbf{x},\eta \right)=-4\Lambda_{ij}^{lm} \mathcal{S}_{lm}^{(2)} \left( \mathbf{x},\eta \right) \ ,
\end{eqnarray}
where the source term is
\begin{eqnarray}\label{eq:ehss}
 \mathcal{S}_{lm}^{(2)} \left( \mathbf{x},\eta \right)&=& \partial_l\phi^{(1)}\partial_m\phi^{(1)}-\frac{4}{3(1+w)}\partial_l\phi^{(1)}\partial_m\phi^{(1)}-\partial_l\psi^{(1)}\partial_m\phi^{(1)}-\frac{4}{3(1+w)\mathcal{H}}\partial_l\psi^{(1)'}\partial_m\phi^{(1)} \nonumber\\
 &-&\partial_l\phi^{(1)}\partial_m\psi^{(1)}+3\partial_l\psi^{(1)}\partial_m\psi^{(1)}-\frac{4}{3(1+w)\mathcal{H}}\partial_l\phi^{(1)}\partial_m\psi^{(1)'}\nonumber\\
 &-&\frac{4}{3(1+w)\mathcal{H}^2}\partial_l\psi^{(1)'}\partial_m\psi^{(1)'}+2\phi^{(1)}\partial_m\partial_l\phi^{(1)}+2\psi^{(1)}\partial_l \partial_m \psi^{(1)} \ .
\end{eqnarray}
Eq.~(\ref{eq:eh}) is the equation of motion of second order SIGWs. In Eq.~(\ref{eq:eh})--Eq.~(\ref{eq:ehss}), we have used Eq.~(\ref{eq:r1}) and Eq.~(\ref{eq:u1}) to represent first order density perturbation $\rho^{(1)}$ and first order velocity perturbation $u_i^{(1)}$ as functions of first order scalar perturbations. As shown in Eq.~(\ref{eq:eh})--Eq.~(\ref{eq:ehss}), in the case of a perfect fluid, second order SIGWs depend only on first order scalar perturbations and physical quantities related to the \ac{FLRW} background spacetime: $w$ and $\mathcal{H}$. If there is an anisotropic stress tensor $\Pi_{ij}^{(2)}$, then Eq.~(\ref{eq:eh}) can be rewritten as
\begin{eqnarray}
     h_{ij}^{(2)''}\left(\mathbf{x},\eta \right)+2\mathcal{H}h_{ij}^{(2)'}\left(\mathbf{x},\eta \right)-\Delta h_{ij}^{(2)}\left(\mathbf{x},\eta \right)=-4\Lambda_{ij}^{lm} \mathcal{S}_{lm} \left( \mathbf{x},\eta \right)+\sigma_{ij}^{\mathrm{TT},(2)} \ ,
\end{eqnarray}
where $\sigma_{ij}^{\mathrm{TT},(2)}$ is the transverse and traceless part of the three-dimensional anisotropic stress tensor $\Pi_{ij}^{(2)}$. For an anisotropic stress tensor $\Pi_{ij}=0$, during the \ac{RD} era, the source term $S_{lm}$ is given by
\begin{eqnarray}\label{eq:S2}
		\mathcal{S}^{(2)}_{lm}(\mathbf{x},\eta)&&= \partial_{l} \phi^{(1)} \partial_{m} \phi^{(1)}+4 \phi^{(1)} \partial_{l} \partial_{m} \phi^{(1)}
		-\frac{1}{ \mathcal{H}}\left(\partial_{l} \phi^{(1)'} \partial_{m} \phi^{(1)}+\partial_{l} \phi^{(1)} \partial_{m} \phi^{(1)'}\right)\nonumber\\
  &&-\frac{1}{ \mathcal{H}^{2}} \partial_{l}  \phi^{(1)'} \partial_{m}  \phi^{(1)'} \ ,
\end{eqnarray}
where we have used the relation: $\psi^{(1)}=\phi^{(1)}$ in Eq.~(\ref{eq:phi1}). The Fourier components of $h_{ij}^{(2)}(\mathbf{x},\eta)$ in terms of the polarization tensors $\varepsilon_{i j}^{\lambda}(\mathbf{k})$ $(\lambda=+,\times)$ are defined as
\begin{equation}\label{eq:h}
\begin{aligned}
h_{i j}^{(2)}(\mathbf{x}, \eta)=\int \frac{\mathrm{d}^3 k}{(2 \pi)^{3 / 2}} e^{i \mathbf{k} \cdot \mathbf{x}}\left(h_{\mathbf{k}}^{+,(2)}(\eta) \varepsilon_{i j}^{+}(\mathbf{k})+h_{\mathbf{k}}^{\times,(2)}(\eta) \varepsilon_{i j}^{\times}(\mathbf{k})\right) \  .
\end{aligned}
\end{equation}
By solving Eq.~(\ref{eq:eh}) with the source term in Eq.~(\ref{eq:S2}), we obtain the explicit expression of $h_{\mathbf{k}}^{\lambda,(2)}(\eta)$, namely
\begin{eqnarray}\label{eq:2h0} 
  	h^{\lambda,(2)}_{\mathbf{k}}(\eta)=\int \frac{d^3 p}{(2 \pi)^{3 / 2}}  \varepsilon^{\lambda, l m}(\mathbf{k})p_l p_m  I_{h}^{(2)}(|\mathbf{k}-\mathbf{p}|,|\mathbf{p}|,\eta)\zeta_{\mathbf{k}-\mathbf{p}}  \zeta_{\mathbf{p}}  \ , 
\end{eqnarray}
where $I^{(2)}_h$ is the second order kernel function. The analytical expression of second order kernel function can be found in Refs.~\cite{Kohri:2018awv}. 

\section{Scalar induced gravitational waves in $f(R)$ gravity}\label{sec:3}
In this section, we investigate the second order SIGWs in $f(R)$ gravity theory.  We show that the effects of modified gravity in $f(R)$ theory can be characterized by an effective anisotropic stress tensor. The presence of this effective anisotropic stress tensor causes the first order scalar perturbation $\psi^{(1)}$ to differ from $\phi^{(1)}$, and it also influences the equation of motion of the second order SIGWs. The field equation of $f(R)$ modified gravity is given by \cite{Olmo:2011uz}
\begin{eqnarray}\label{eq:ef}
    F R_{\mu \nu}-\frac{1}{2} g_{\mu \nu} f+\left(g_{\mu \nu} \square-\nabla_\mu \nabla_\nu \right) F=8 \pi G T^{M}_{\mu \nu} \ ,
\end{eqnarray}
where we have set $F \equiv f_R= \mathrm{d} f(R) / \mathrm{d} R$. To investigate the second order SIGWs in $f(R)$ gravity theory, a straightforward idea is to substitute the metric perturbation in Eq.~(\ref{eq:ds}) directly into Eq.~(\ref{eq:ef}) and simplify the perturbation equations at each order. However, this approach would make the calculation process extremely complex and lengthy, which is not conducive to our study of the second order SIGWs in $f(R)$ gravity theory under general conditions. As shown in Sec.\ref{sec:2.1}, we can represent the density perturbation $\rho^{(n)}$ and velocity perturbation $u_{i}^{(n)}$ as functions of scalar perturbations by perturbing the time-time component and time-space component of the Einstein field equation. Furthermore, the pressure perturbations $p^{(n)}$ can be directly expressed as a density perturbation through the definition of the speed of sound $c^2_s=p^{(1)}/\rho^{(1)}$ and the equation of state $w=p^{(0)}/\rho^{(0)}$. Therefore, in the case of a perfect fluid, the density perturbation $\rho^{(n)}$, pressure perturbation $p^{(n)}$, and velocity perturbation $u_i^{(n)}$ do not explicitly appear in the equations of motion  of cosmological perturbations. We only need to know the parameters $w$ and $c_s^2$ to determine the equations of motion  of cosmological perturbations.

Similarly, to avoid unnecessary complex calculations, we investigate the second order SIGWs in $f(R)$ gravity theory using  described in Refs.~\cite{DeFelice:2010aj,Arjona:2019rfn,Arjona:2020gtm,Cardona:2022lcz}, where the modified gravity effect on the Einstein field equation is redefined in an equivalent energy-momentum tensor of a non-perfect fluid $T_{\mu\nu}^{f(R)}$. Since the physical quantities of the perfect fluid part in the equivalent energy-momentum tensor $T_{\mu\nu}^{f(R)}$ can be formally expressed as functions of parameters $w$, $c_s$, and the first order scalar perturbations: $\psi^{(1)}$ and $\phi^{(1)}$, we only need to focus on the impact of the anisotropic stress tensor $\Pi_{ij}^{(n)}$ in $T_{\mu\nu}^{f(R)}$.

\subsection{Effective fluid approach}
We briefly review the main results of the \ac{EFA} in $f(R)$ gravity. The equation of motion of $f(R)$ gravity can be rewritten as the usual Einstein field equation plus an effective fluid by adding and subtracting the Einstein tensor $G_{\mu\nu}$ on the left-hand side of Eq.~(\ref{eq:ef}) and moving everything except the Einstein tensor $G_{\mu\nu}$ to the right-hand side, namely \cite{DeFelice:2010aj}
\begin{eqnarray}\label{eq:efe}
    G_{\mu \nu}=\kappa \left( T^M_{\mu\nu}+T_{\mu\nu}^{f(R)} \right) \ ,
\end{eqnarray}
where $T^M_{\mu\nu}$ is the energy-momentum tensor of the perfect fluid that corresponds to the usual matter. The energy-momentum tensor of the effective fluid is given by
\begin{eqnarray}\label{eq:efT}
    \kappa T_{\mu \nu}^{f(R)} &=&(1-F) R_{\mu \nu}+\frac{1}{2}g_{\mu \nu}(f-R)  -\left(g_{\mu \nu} \square-\nabla_\mu \nabla_\nu\right) F \ .
\end{eqnarray}
By substituting the \ac{FLRW} background metric in Eq.~(\ref{eq:ds0}) into Eq.~(\ref{eq:efe}) in the \ac{EFA}, we obtain the same form of background evolution equation as in general relativity
\begin{eqnarray}\label{eq:111}
    \mathcal{H}^2 &=&\frac{\kappa}{3} a^2\left(\rho^{(0)}_m+\rho^{(0)}_{f}\right) \ , \\
\mathcal{H}' &=&-\frac{\kappa}{6} a^2\left(\left(\rho^{(0)}_m+\rho^{(0)}_{f}\right)+3\left(p^{(0)}_m+ p^{(0)}_{f}\right)\right) \ ,
\end{eqnarray}
where the subscripts $m$ and $f$ denote the usual matter and $f(R)$ theory, respectively. The effective density $\rho^{(0)}_{f}$ and pressure $p^{(0)}_{f}$ in Eq.~(\ref{eq:111}) are given by \cite{Arjona:2018jhh}
\begin{eqnarray}
\kappa \rho^{(0)}_{f} &=&-\frac{f}{2}+3 \mathcal{H}^2 / a^2-3 \mathcal{H} F'/ a^2+3 F \mathcal{H}' / a^2  \ , \\
    \kappa p^{(0)}_{f} &=&\frac{f}{2}-\mathcal{H}^2 / a^2-2 F \mathcal{H}^2 / a^2+\mathcal{H} F' / a^2 -2 \mathcal{H}' / a^2-F \mathcal{H}' / a^2+F'' / a^2 \ .
\end{eqnarray}
If we set $\rho^{(0)}_{\mathrm{tot}}=\rho^{(0)}_m+\rho^{(0)}_{f}$ and $p^{(0)}_{\mathrm{tot}}=p^{(0)}_m+ p^{(0)}_{f}$, then the background evolution equation in $f(R)$ gravity will have the same structure as that in general relativity. 

\subsection{First order scalar perturbations}
We now consider the  first order perturbation of the effective fluid energy-momentum tensor in $f(R)$ gravity, represented as 
\begin{eqnarray}\label{eq:fT1}
    \left(T^{f,(1)}\right)^{~0}_{0}&=&-\rho^{(1)}_{f}  \ , \
\left(T^{f,(1)}\right)_{0}^{~j}=-\left(\rho^{(0)}_{f}+p^{(0)}_{f} \right)u^{(1),j}_f \ , \nonumber\\
\left(T^{f,(1)}\right)^{~j}_{i}&=&\delta_i^{~j} ~p^{(1)}_{f} +p^{(0)}_{f}\left(\Pi^{f,(1)}\right)^{~j}_{i} \ , 
\end{eqnarray}
where the symbol $\left(T^{f,(1)}\right)^{~\nu}_{\mu}$ represents the $\mu-\nu$ component of the first order perturbation of the effective fluid energy-momentum tensor. By calculating the first order perturbation of Eq.~(\ref{eq:efT}) and comparing it with Eq.~(\ref{eq:fT1}), we obtain
\begin{eqnarray}
   \rho^{(1)}_{f}&&=-\frac{1}{\kappa a^2 }\left(\frac{a^2}{2}f^{(1)}+6\mathcal{H}^2\phi^{(1)}-3\mathcal{H}'\left(F^{(1)}-2F^{(0)}\phi^{(1)}\right)-3F^{(0)'}\psi^{(1)'} \right. \nonumber\\
   &&\left.+3\mathcal{H}\left( F^{(1)'}-2F^{(0)'}\phi^{(1)}+F^{(0)}\phi^{(1)'}+\left(2+F^{(0)} \right)\psi^{(1)'} \right) -\Delta F^{(1)} \right. \nonumber\\
    && \left. +F^{(0)}\left( 3\psi^{(1)''}+\Delta\phi^{(1)} \right)-2\Delta \psi^{(1)} \right) \ ,   \label{eq:1rf}\\
     p^{(1)}_{f}&&=\frac{1}{2\kappa a^2}\left(2F^{(1)''}+a^2f^{(1)} +2\mathcal{H}'\left(-F^{(1)} +2\left(2+F^{(0)}  \right)\phi^{(1)}  \right) +4\mathcal{H}^2\left(-F^{(1)} \right.\right. \nonumber\\
     && \left.\left.+\left(1+2F^{(0)}   \right)\phi^{(1)}   \right)+2\mathcal{H}\left(F^{(1)'}-2F^{(0)'}\phi^{(1)} +\left( 2+F^{(0)} \right)\phi^{(1)'} \right.\right. \nonumber\\
     && \left.\left. +\left( 4+5F^{(0)} \right) \psi^{(1)'}   \right)  +2\left( -2F^{(0)''}\phi^{(1)}-F^{(0)'}\left( \phi^{(1)'}+2\psi^{(1)'} \right) \right.\right. \nonumber\\
     && \left.\left. +\left(2+F^{(0)} \right)\psi^{(1)''} -\Delta F^{(1)}+\Delta \phi^{(1)}-\left(1+F^{(0)} \right)\Delta \psi^{(1)} \right) 
 \right) \ ,  \label{eq:1pf} \\
     \left(u^{f,(1)}\right)^{j}&&=-\frac{1}{\kappa a^2\left(\rho^{(0)}_{f}+p^{(0)}_{f} \right)}\left( \partial^j F^{(1)'}-F^{(0)'}\partial^j\phi^{(1)}+\mathcal{H}\left(-\partial^jF^{(1)}\right. \right.\nonumber\\
     &&\left.\left.-2\left(-1+F^{(0)} \right)\partial^j \phi^{(1)}  \right) -2\left(-1+F^{(0)} \right)\partial^j\psi^{(1)'} \right) \ ,  \label{eq:1uf} \\
     \left(\Pi^{f,(1)}\right)^{~j}_{i}&&=\frac{1}{\kappa a^2 p^{(0)}_{f}}\left( \partial_i\partial^j F^{(1)}+\left( F^{(0)}-1 \right)\left(\partial_i\partial^j \phi^{(1)}-\partial_i\partial^j \psi^{(1)} \right)  \right) \label{eq:1Pif} \ ,
\end{eqnarray}
where $F^{(0)}$ denotes the background value of $F \equiv \mathrm{d} f(R) / \mathrm{d} R$. $f^{(1)}$ and $F^{(1)}$ represent the first order perturbation of $f(R)$ and $F(R)$, respectively. In the context of $f(R)$ modified gravity theory, if we set: $\rho^{(1)}_{\mathrm{tot}}=\rho^{(1)}_m+\rho^{(1)}_{f}$,  $p^{(1)}_{\mathrm{tot}}=p^{(1)}_m+ p^{(1)}_{f}$, $u^{(1),i}_{\mathrm{tot}}=u^{(1),i}_m+ u^{(1),i}_{f}$, then the first order cosmological perturbation equation in $f(R)$ theory will have a similar form to the first order perturbation equation in general relativity. Furthermore, the first order cosmological perturbation equation in $f(R)$ theory will exhibit a non-zero effective anisotropic stress tensor $\left(\Pi^{f,(1)}\right)_{ij}$, which is absent in the first order perturbation equation of general relativity. The first order cosmological perturbation equation in $f(R)$ modified gravity theory can be represented as
\begin{eqnarray}
   & &-6\mathcal{H}\psi^{(1)'}+2\Delta\psi^{(1)}=\kappa a^2\left( \rho^{(1)}_{\mathrm{tot}}+2\rho^{(0)}_{\mathrm{tot}}\phi^{(1)}\right)  \ , \label{eq:fG100}\\ 
& &2\left( \mathcal{H} \partial_i \phi^{(1)}+\partial_i \psi^{(1)'} \right)=-\kappa\left(1+w_{\mathrm{tot}} \right)a^2 \rho_{\mathrm{tot}}^{(0)} u^{(1)}_{\mathrm{tot},i} \ , \label{eq:fG10i}\\  
 & &\delta_{ij} \left( 2\mathcal{H}\left( \phi^{(1)}+\psi^{(1)}\right)+4\mathcal{H}'\left(\phi^{(1)}+\psi^{(1)} \right)+2\mathcal{H}\left( \phi^{(1)'}+2\psi^{(1)'} \right)+2\psi^{(1)''}+\Delta\phi^{(1)} \right. \nonumber\\
& &~~\left.-\Delta\psi^{(1)} \right)=\kappa a^2 \delta_{ij} \left( \left(c_{s,\mathrm{tot}}\right)^2\rho_{\mathrm{tot}}^{(1)}-2w_{\mathrm{tot}}\rho^{(0)}_{\mathrm{tot}}\psi^{(1)} \right)+\kappa p^{(0)}_{f}\left(\Pi^{f,(1)}\right)_{ij} \ , \label{eq:fG1ij}
\end{eqnarray}
where we have set $w_{\mathrm{tot}}=p^{(0)}_{\mathrm{tot}}/\rho^{(0)}_{\mathrm{tot}}$, and $\left(c_{s,\mathrm{tot}}\right)^2=p^{(1)}_{\mathrm{tot}}/\rho^{(1)}_{\mathrm{tot}}$. Using the decomposition operators, we can derive the equations of motion of first order scalar perturbations in 
$f(R)$ modified gravity
\newtcolorbox{mymathbox}[1][]{colback=white, #1}

\begin{mymathbox}[ams align, title=Equations of motion of first order scalar perturbations in $f(R)$ gravity , colframe=white!40!black]
     &2\mathcal{H} \left(3(\left(c_{s,\mathrm{tot}}\right)^2-w_{\mathrm{tot}})\mathcal{H}\phi^{(1)}+\phi^{(1)'}+(2+3\left(c_{s,\mathrm{tot}}\right)^2)\psi^{(1)'}   \right)  +2\psi^{(1)''}+\Delta \phi^{(1)}  \label{eq:fs1} \nonumber\\
     &-\Delta \psi^{(1)}-2\left(c_{s,\mathrm{tot}}\right)^2\Delta \psi^{(1)}=0 \ , 
      \\ \nonumber\\
     &\partial_i\partial_j\psi^{(1)}-\partial_i\partial_j\phi^{(1)}= \partial_i\partial_j F^{(1)}+\left( F^{(0)}-1 \right)\left(\partial_i\partial_j \phi^{(1)}-\partial_i\partial_j \psi^{(1)} \right) \ . \label{eq:fs2}
\end{mymathbox}
Eq.~(\ref{eq:fs1}) and Eq.~(\ref{eq:fs2}) are similar to the equations of motion of first order scalar perturbations in general relativity (Eq.~(\ref{eq:s1}) and Eq.~(\ref{eq:s2})), except for the contribution of the non-zero effective anisotropic stress tensor. It is worth noting that we have not made any approximations to the effective fluid energy-momentum tensor $T^{f(R)}_{\mu\nu}$ in Eq.~(\ref{eq:efT}). We express the perturbation equations in $f(R)$ gravity in a form similar to that in general relativity, allowing us to make direct comparisons the perturbation equation in general relativity. The effects of the effective fluid energy-momentum tensor $T^{f(R)}_{\mu\nu}$ are encoded in the parameters $w_{\mathrm{tot}}$ and $c_{s,\mathrm{tot}}$, as well as in the effective anisotropic stress tensor $\Pi^{f,(1)}_{ij}$. By simplifying Eq.~(\ref{eq:fs2}), we  obtain
\begin{eqnarray}\label{eq:F1/F0}
    \psi^{(1)}-\phi^{(1)}=\frac{F^{(1)}}{F^{(0)}} \ ,
\end{eqnarray}
where $F^{(1)}=F_{R}^{(0)}R^{(1)}$. Eq.~(\ref{eq:F1/F0}) represents the effect of the effective anisotropic stress tensor on the first order scalar perturbation in $f(R)$ modified gravity theory.  In general relativity, $F_{R}^{(0)}=0$ leads to $\phi^{(1)}=\psi^{(1)}$, which is consistent with the result in Eq.~(\ref{eq:s2}).

\subsection{Scalar induced gravitational waves}
We now turn our attention to the second order SIGWs in $f(R)$ modified gravity theory. Similar to the first order scalar perturbation, the second order perturbations of the effective fluid energy-momentum tensor $T^{f(R)}_{\mu\nu}$ will generate a non-zero anisotropic stress tensor. Specifically, substituting Eq.~(\ref{eq:ds}) into Eq.~(\ref{eq:efe}) and retaining the transverse and traceless part of the second order perturbation yields the equation of motion of the second order SIGWs in $f(R)$ modified gravity theory
\begin{eqnarray}\label{eq:ehf}
    h_{ij}^{(2)''}\left(\mathbf{x},\eta \right)+2\mathcal{H}h_{ij}^{(2)'}\left(\mathbf{x},\eta \right)-\Delta h_{ij}^{(2)}\left(\mathbf{x},\eta \right)=-\frac{4}{F^{(0)}}\Lambda_{ij}^{lm} \left(\mathcal{S}_{lm}^{(2)} \left( \mathbf{x},\eta \right)+\Pi^{(2)}_{lm}\left( \mathbf{x},\eta \right)\right) \ ,
\end{eqnarray}
where 
\begin{eqnarray}\label{eq:pitt}
   \Pi^{(2)}_{lm}&=&-\frac{1}{4} \left(-h_{ij}^{(2)''}\left( -1+F^{(0)}\right)-h_{ij}^{(2)'}\left(2\mathcal{H}\left(-1+F^{(0)}\right)+F^{(0)'} \right)+\Delta h_{ij}^{(2)}\left(-1+F^{(0)} \right)  \right) \nonumber\\
    &+&\left(3F^{(1)}+\left(-1+F^{(0)}  \right)\left(\phi^{(1)}-\psi ^{(1)} \right)  \right)\partial_i\partial_j\psi^{(1)} \nonumber\\
    &-&\left( F^{(1)}-\left( -1+F^{(0)} \right)\left(\psi^{(1)} +\phi^{(1)} \right) \right) \partial_i\partial_j\phi^{(1)}  \ .
\end{eqnarray}
In Eq.~(\ref{eq:pitt}), we have neglected terms that are not transverse and traceless.  By simplifying Eq.~(\ref{eq:ehf}), we can obtain the equations of motion of second order SIGWs for $f(R)$ modified gravity theory in momentum space

\begin{mymathbox}[ams align, title=Equations of motion of second order SIGWs in $f(R)$ gravity, colframe=white!40!black]
\label{eq:ehfR}
    h_{\mathbf{k}}^{\lambda,(2)''}\left(\eta \right)+\frac{1}{F^{(0)}}\left(2\mathcal{H}F^{(0)} +F^{(0)'} \right)h_{\mathbf{k}}^{\lambda,(2)'}\left(\eta \right)&+k^2 h_{\mathbf{k}}^{\lambda,(2)}\left(\eta \right) \nonumber\\
    &=\frac{4}{F^{(0)}} \left(\mathcal{S}_{\mathbf{k}}^{\lambda,(2)} \left(\eta \right)+\sigma^{\lambda,(2)}_{\mathbf{k}}\left( \eta \right)\right) \ ,
\end{mymathbox}

where 

\begin{mymathbox}[ams align, title=Source terms of SIGWs in $f(R)$ gravity, colframe=white!40!black]
    \mathcal{S}_{\mathbf{k}}^{\lambda,(2)}(\eta)&=\int\frac{d^3q}{(2\pi)^{3/2}}\varepsilon^{\lambda, lm}(\mathbf{k})p_lp_m\left(\left(1+\frac{4}{3(1+w_{\mathrm{tot}})}\right)\phi^{(1)}_{\mathbf{k}-\mathbf{p}}(\eta) \phi^{(1)}_{\mathbf{p}}(\eta)+2\psi^{(1)}_{\mathbf{k}-\mathbf{p}}(\eta) \phi^{(1)}_{\mathbf{p}}(\eta) \right. \nonumber\\
    &\left.+\frac{8}{3(1+w_{\mathrm{tot}})\mathcal{H}}\psi^{(1)'}_{\mathbf{k}-\mathbf{p}}(\eta) \phi^{(1)}_{\mathbf{p}}(\eta)+\frac{4}{3(1+w_{\mathrm{tot}})\mathcal{H}^2}\psi^{(1)'}_{\mathbf{k}-\mathbf{p}}(\eta) \psi^{(1)'}_{\mathbf{p}}(\eta)  \right. \nonumber\\
    &\left. -\psi^{(1)}_{\mathbf{k}-\mathbf{p}}(\eta) \psi^{(1)}_{\mathbf{p}}(\eta) \right) \ , \label{eq:Ss}\\
 \sigma_{\mathbf{k}}^{\lambda,(2)}(\eta)&=\int\frac{d^3q}{(2\pi)^{3/2}}\varepsilon^{\lambda, lm}(\mathbf{k})p_lp_m\left(-\left( F^{(1)}_{\mathbf{k}-\mathbf{p}}-\left( -1+F^{(0)} \right)\left(\psi^{(1)}_{\mathbf{k}-\mathbf{p}} +\phi^{(1)}_{\mathbf{k}-\mathbf{p}} \right) \right) \phi^{(1)}\right.\nonumber\\
 &\left.+\left(3F^{(1)}_{\mathbf{k}-\mathbf{p}}+\left(-1+F^{(0)}  \right)\left(\phi^{(1)}_{\mathbf{k}-\mathbf{p}}-\psi ^{(1)}_{\mathbf{k}-\mathbf{p}} \right)  \right)\psi^{(1)} \right) \ . \label{eq:Ssigma}
\end{mymathbox}

The symbols $\lambda=+$, and $\times$ in Eq.~(\ref{eq:ehfR}) represent the two polarization modes of gravitational waves. Eq.~(\ref{eq:ehfR})--Eq.~(\ref{eq:Ssigma}) provide the explicit expression of the equation of motion of second order SIGWs in $f(R)$ modified gravity theory. Furthermore, an additional propagating degree of freedom, the scalaron field $\phi_{s}$, is a distinctive feature of the richer structure of $f(R)$ gravity \cite{Starobinsky:1980te,Katsuragawa:2019uto,Myung:2016zdl,Gong:2017bru,Moretti:2019yhs,Vainio:2016qas}. Its equation can be obtained by taking the trace of Eq.~(\ref{eq:ef}), which produces
\begin{eqnarray}\label{eq:sme0}
    \square \phi_{s}=\frac{dV_{\mathrm{eff}}}{d\phi_{s}} \ \ , \  \  \frac{dV_{\mathrm{eff}}}{d\phi_{s}}\equiv\frac{1}{3}\left( 2f-\phi_{s} R +\kappa T^M \right) \ ,
\end{eqnarray}
where $\phi_{s}\equiv F(R)$. $T^M$ is the  trace of energy-momentum tensor. First order perturbation of Eq.~(\ref{eq:sme0}) gives
\begin{eqnarray}\label{eq:sme1.0}
    \square^{(0)} \phi_{s}^{(1)}+\square^{(1)} \phi_{s}^{(0)}&=&\frac{1}{3}\left(  2f^{(1)}-F^{(0)}R^{(1)}-F^{(1)}R^{(0)}+\kappa T^{M,(1)} \right) \nonumber\\
    &=&\frac{1}{3}\left( \frac{F^{(0)}}{F^{(0)}_R}-R^{(0)}  \right)F^{(0)}_{R}R^{(1)}+\frac{\kappa}{3} T^{M,(1)} \ ,
\end{eqnarray}
where $\phi^{(0)}_{s}= F^{(0)}$, and $\phi^{(1)}_{s}=F^{(1)}= F_{R}^{(0)}R^{(1)}$. The mass of scalaron field is given by
\begin{eqnarray}
    m_s^2\equiv\frac{d^2V_{\mathrm{eff}}}{d\phi_{s}^2}=\frac{1}{3}\left( \frac{F^{(0)}}{F^{(0)}_R}-R^{(0)}  \right) \ .
\end{eqnarray}
Then, Eq.~(\ref{eq:sme1.0}) on \ac{FLRW} space-time can be rewritten as \cite{Katsuragawa:2019uto}
\begin{eqnarray}\label{eq:phis1}
   \phi_{s}^{(1)''}+2\mathcal{H}\phi_{s}^{(1)'}-\left(\Delta+m^2_s  \right)\phi_{s}^{(1)}=-\square^{(1)} \phi_{s}^{(0)} +\frac{\kappa}{3} T^{M,(1)} \ .
\end{eqnarray}
The scalaron field $\phi_{s}$ provides an additional massive scalar mode for gravitational waves with source term: $\mathcal{S}^{(1)}_{s}=-\square^{(1)} \phi_{s}^{(0)} +\frac{\kappa}{3} T^{M,(1)}$. Similar to the calculation in general relativity, the solution to Eq.~(\ref{eq:ehfR}) and  Eq.~(\ref{eq:phis1}) in momentum space can be expressed as 
\begin{eqnarray}
    h_{\mathbf{k}}^{\lambda,(2)}\left(\eta \right)&=&\int\frac{d^3q}{(2\pi)^{3/2}}\varepsilon^{\lambda, lm}(\mathbf{k})p_lp_m I^{(2)}_{hf} \left(u,v,x \right) \zeta_{\mathbf{k}-\mathbf{p}}\zeta_{\mathbf{p}} \ , \label{eq:sol1}\\
     \phi_{s,\mathbf{k}}^{(1)}\left(\eta \right)&\equiv&T_{s}(x)\zeta_{\mathbf{k}}=-\frac{2F_R^{(0)}}{a^2}\left(6\left(\mathcal{H}'+\mathcal{H}^2 \right) T_{\phi}(x)+3\mathcal{H}k\left( \frac{d}{dx}T_{\phi}(x) +3\frac{d}{dx}T_{\psi}(x)  \right) \right. \nonumber\\
     &&\left.+3k^2\frac{d^2}{d^2x}T_{\psi}(x)-k^2T_{\phi}(x)+2k^2T_{\psi}(x)   \right)\zeta_{\mathbf{k}} \label{eq:sol2}\ ,
\end{eqnarray}
where we have defined $x=\eta |\mathbf{k}|$, $|\mathbf{k}-\mathbf{p}|=u|\mathbf{k}|$, and $|\mathbf{p}|=v|\mathbf{k}|$. The first order scalar perturbations: $\phi^{(1)}$, $\psi^{(1)}$ and the first order perturbation of scalar mode $\phi_s^{(1)}$ in Eq.~(\ref{eq:sol2}) have been written as
\begin{eqnarray}
    \phi^{(1)}_{\mathbf{k}}(\eta)&=&\left(\frac{3+3w_{\mathrm{tot}}}{5+3w_{\mathrm{tot}}}\right) T_{\phi}(x)\zeta_{\mathbf{k}} \ \  , \  \  \psi^{(1)}_{\mathbf{k}}(\eta)=\left(\frac{3+3w_{\mathrm{tot}}}{5+3w_{\mathrm{tot}}}\right) T_{\phi}(x)\zeta_{\mathbf{k}} \ , \\ \nonumber\\
    \phi_{s,\mathbf{k}}^{(1)}\left(\eta \right)&=&F^{(1)}=F^{(0)}_{R}R^{(1)}=T_{s}(x)\zeta_{\mathbf{k}} \ .
\end{eqnarray}
The second order kernel function $I^{(2)}_{hf}(u,v,x)$ in Eq.~(\ref{eq:sol1}) satisfies
\begin{eqnarray}\label{eq:kernel}
    I^{(2)''}_{hf}\left(u,v,x  \right)&+&\frac{1}{F^{(0)}}\left(2\mathcal{H}F^{(0)}+F^{(0)'} \right)I^{(2)'}_{hf}\left(u,v,x  \right)\nonumber\\
    &+&k^2I^{(2)}_{hf}\left(u,v,x  \right)=4\left(f^{(2)}_s\left(u,v,x  \right)+f^{(2)}_{\sigma}\left(u,v,x  \right) \right) \ ,
\end{eqnarray}
where
    \begin{align}
        f^{(2)}_s\left(u,v,x  \right)&=\left(1+\frac{4}{3(1+w_{\mathrm{tot}})}\right)T_{\phi}(ux) T_{\psi}(vx)+2T_{\psi}(ux) T_{\phi}(vx) -T_{\psi}(ux) T_{\psi}(vx) \nonumber\\
    &+\frac{8u}{3(1+w_{\mathrm{tot}})\mathcal{H}}\frac{d}{d(ux)}T_{\psi}(ux) T_{\phi}(vx)   +\frac{4uv}{3(1+w_{\mathrm{tot}})\mathcal{H}^2}\frac{d}{d(ux)}T_{\psi}(ux)\frac{d}{d(vx)}T_{\psi}(vx)           \ ,  \\
    f^{(2)}_{\sigma}\left(u,v,x  \right)&=2T_{\psi}(ux) T_{s}(vx)+ \left(F^{(0)}-1 \right)\left(3T_{\psi}(ux) T_{\psi}(vx) -T_{\phi}(ux)T_{\phi}(vx) \right)              \ . 
    \end{align}
As shown in Eq.~(\ref{eq:kernel}), the dynamical and source terms of the kernel function $I^{(2)}_{hf}(u,v,x)$ for the second order SIGWs are also distinct. Furthermore, in contrast to the second order SIGWs in general relativity, $f(R)$ modified gravity theory includes an extra scalar mode $\phi_s$.  

In this section, we investigate the equations of motion for second-order SIGWs and first-order scalar mode within $f(R)$ theory and offer the corresponding formal solutions. It is crucial to note that Eq.~(\ref{eq:fs1}), Eq.~(\ref{eq:fs2}), Eq.~(\ref{eq:phis1}),   and Eq.~(\ref{eq:kernel}) are independent of any specific theoretical model; they are applicable to all forms of $f(R)$  theory during all cosmological dominant eras. For a given specific form of $f(R)$ theory, we can determine the exact form of $F^{(0)}$ and derive the corresponding explicit expression for second-order SIGWs.

\section{Energy density spectra and PTA observation}\label{sec:4}
In this section, we present the explicit expressions of the energy density spectra of SIGWs in $f(R)$ theory. We compute the energy density spectrum for the HS model. By integrating the current PTA observational data, we can constrain the parameter space of the HS model. 

\subsection{Energy density spectra}
The power spectrum of \acp{GW} is defined as
\begin{equation}\label{eq:Ph}
\left\langle h_{\mathbf{k}}^{\lambda}(\eta) h_{\mathbf{k}^{\prime}}^{\lambda'}(\eta)\right\rangle \equiv \delta^{(3)}\left(\mathbf{k}+\mathbf{k}'\right) \delta^{\lambda \lambda'} \frac{2 \pi^2}{k^3} \mathcal{P}_h(\eta, k) \  ,
\end{equation}
where $\lambda=\times, +$ and $s$. Substituting Eq.~(\ref{eq:sol1}) into Eq.~(\ref{eq:Ph}), we obtain the explicit expression of the power spectrum  for the $\times$ and $+$ polarization states
\begin{equation}\label{eq:Php}
\mathcal{P}_h^{(\times) \text {or }(+)}(\eta, k)=4 \int_0^{\infty} \mathrm{d} v \int_{|1-v|}^{1+v} \mathrm{~d} u\left[\frac{4 v^2-\left(1+v^2-u^2\right)^2}{4 u v}\right]^2 \left(I^{(2)}_{hf}(u, v, x) \right)^2\mathcal{P}_{\zeta}(k v) \mathcal{P}_{\zeta}(k u) \ ,
\end{equation}
where $\mathcal{P}_{\zeta}\left(k \right)$ is the power spectrum of primordial curvature perturbation. The second order kernel function $I^{(2)}_{hf}(u, v, x)$ is provided in Eq.~(\ref{eq:kernel}). In $f(R)$ theory, the second order kernel function $I^{(2)}_{hf}$ for the $+$ and $\times$ modes are distinct from the results obtained in general relativity. For the first order scalar mode in Eq.~(\ref{eq:sol2}), the corresponding power spectrum can be written as $\mathcal{P}_s\left(k,\eta \right)=\mathcal{P}_{\zeta}\left( k\right)T^2_{s}\left(x \right)$, where the first order transfer function $T_{s}\left( x\right)$ is defined in Eq.~(\ref{eq:sol2}). The energy density spectrum of SIGWs in $f(R)$ theory during the \ac{RD} era is expressed as
\begin{equation}
    \overline{\Omega}_{\mathrm{GW}}(\eta, k) = \overline{\frac{\rho_{\mathrm{GW}}(\eta,k)}{\rho_{\mathrm{tot}}(\eta)}} = \overline{\frac{x^2}{6}\mathcal{P}_h(\eta,k)}  \ ,
\end{equation}
where
\begin{equation}
    \mathcal{P}_h(\eta,k) = \frac{1}{4}\mathcal{P}^{(2)}_h(\eta,k) +\mathcal{P}_s(\eta,k) \  ,
\end{equation}
is the total power spectrum of SIGWs in $f(R)$ theory. The current energy density spectrum of second order SIGWs is given by \cite{Wang:2019kaf}
\begin{equation}
    \bar{\Omega}_{\mathrm{GW,0}}(k) = \Omega_{\mathrm{rad},0}\left(\frac{g_{*,\rho,e}}{g_{*,\rho,0}}\right)\left(\frac{g_{*,s,0}}{g_{*,s,e}}\right)^{4/3}\bar{\Omega}_{\mathrm{GW}}(\eta,k) \ ,
\end{equation}
where $\bar{\Omega}_{\mathrm{GW}}(\eta,k)$ represents the energy density spectrum of second order SIGWs during the \ac{RD} era, while $\bar{\Omega}_{\mathrm{GW,0}}(k)$ denotes the current energy density spectrum of SIGWs. The effect numbers of relativistic species $g_{*,\rho}$ and $g_{*,s}$ can be found in Ref.~\cite{Saikawa:2018rcs}. $\Omega_{\mathrm{rad},0}$ ($ =4.2\times 10^{-5}h^{-2}$) is the energy density fraction of radiations today, and the dimensionless Hubble constant is $h = 0.6736$ \cite{Planck:2018vyg}.

To illustrate in detail how current PTA observational data constrain the $f(R)$ modified theory via SIGWs, we consider the HS model in the $f(R)$ modified gravity theory to calculate the energy density spectrum of SIGWs. More precisely, the HS model is defined as \cite{Hu:2007nk}
\begin{equation}\label{eq:HS}
   f(R)=R-m^2 \frac{c_1\left(R / m^2\right)^n}{1+c_2\left(R / m^2\right)^n} 
\ , 
\end{equation}
where $c_1, c_2$ are two free parameters. $m$ and $n$ are positive constants with $n$ usually taking positive integer values i.e., $n=1,2, \cdots$. In the rest of our paper we set $n=1$ \cite{Arjona:2018jhh}.  We consider the SIGWs during the \ac{RD} era, where $ w_{\mathrm{tot}} \approx  \frac{1}{3}$ and $\left(c_{s,\mathrm{tot}}\right)^2 \approx  \frac{1}{3}$. In this scenario, we have assumed that the energy-momentum tensor provided by matter is significantly greater than the modified gravity effects, specifically, $w_{\mathrm{tot}}=p^{(0)}_{\mathrm{tot}}/\rho^{(0)}_{\mathrm{tot}}\approx p^{(0)}_{m}/\rho^{(0)}_{m}$, and $\left(c_{s,\mathrm{tot}}\right)^2=p^{(1)}_{\mathrm{tot}}/\rho^{(1)}_{\mathrm{tot}}\approx p^{(1)}_{m}/\rho^{(1)}_{m}$. Then, the conformal Hubble parameter $\mathcal{H}$ satisfies $\mathcal{H}^2 = -\mathcal{H}'$, and the Ricci scalar and its first order perturbation satisfy $R^{(0)}=R^{(1)}=0$. At this stage, $F^{(0)} = 1 - c_1$, and the scalar mode $\phi_s = F^{(1)} = 0$. During the \ac{RD} era, the equation of motion for the second order SIGWs depends only on the parameter $c_1$ in the HS model and is independent of other parameters in the model. By substituting Eq.~(\ref{eq:HS}) into Eq.~(\ref{eq:fs1})--Eq.~(\ref{eq:fs2}) and Eq.~(\ref{eq:ehfR}), we can solve the perturbation equations in HS model order by order.

In the above discussion, we used the parameterization method to handle the cosmological perturbation equations in $f(R)$ theory, incorporating part of the modified gravity effects into the parameters $w_{\mathrm{tot}}$ and $c_{s,\mathrm{tot}}$. This method can also be applied to other cosmological dominant era, such as the early matter-dominated era \cite{Assadullahi:2009nf,Alabidi:2013lya,Papanikolaou:2024kjb,Papanikolaou:2022chm}. For SIGWs in $f(R)$ theory during other dominant eras, we need to select different parameters $w_{\mathrm{tot}}$ and $c_{s,\mathrm{tot}}$ and re-solve the cosmological perturbation equations presented in Sec.~\ref{sec:3}. Here, it is crucial to highlight that while the parameterization method is extensively used in various cosmological studies of modified gravity, its application to cosmological perturbations involving modified gravity is conditional. Specifically, we can estimate the  parameters $w_{\mathrm{tot}}$ and $c_{s,\mathrm{tot}}$ and apply the parameterization method to solve the cosmological perturbation equations only when the contribution of energy-momentum tensor provided by matter is significantly exceed the effects of modified gravity. However, if the influence of modified gravity is much greater than that of matter, the parameters $w_{\mathrm{tot}}$ and $c_{s,\mathrm{tot}}$ cannot be predetermined. In this case, we need to rigorously solve equations of motion of cosmological perturbations. 

In the following parts of this section, we calculate the energy density spectrum of second-order SIGWs within the HS model during the \ac{RD} era. We consider a log-normal primordial power spectrum
\begin{equation}
    \mathcal{P}_{\zeta}(k) = \frac{A_\zeta}{\sqrt{2\pi\sigma_*^2}}\exp\left(-\frac{\ln^{2}(k/k_*)}{2\sigma_*^2}\right) \ ,
\end{equation}
where $A_{\zeta}$ is the amplitude of primordial power spectrum and $k_*=2\pi f_*$ is the wavenumber at which the primordial power spectrum has a log-normal peak. 

\begin{figure}[htbp]
    \centering
    \includegraphics[width=.8\columnwidth]{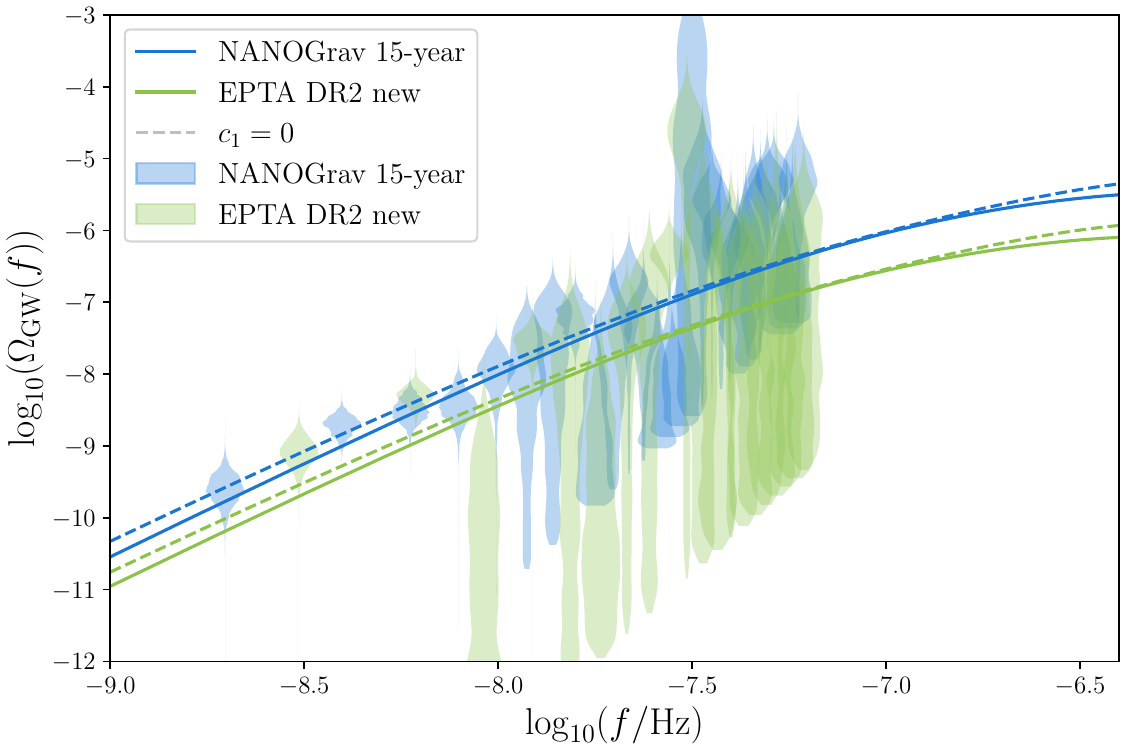}
\caption{\label{fig:violin} The energy density spectra of $f(R)$ modified gravity theory and general relativity. The blue and green solid curves are based on parameters derived from the median values of the posterior distributions of the NANOGrav 15-year and EPTA DR2 new datasets, respectively. 
Specifically, the parameters for the blue solid line are: $\log_{10}(A_{\zeta})=0.35$, $\log_{10}(c_1)=-0.22$, $\log_{10}(f_*/\mathrm{Hz})=-6.11$, and $\sigma_*=1.17$; and for the green solid line, the parameters are: $\log_{10}(A_{\zeta})=0.04$, $\log_{10}(c_1)=-0.26$, $\log_{10}(f_*/\mathrm{Hz})=-6.19$, and $\sigma_*=1.60$. The dashed curves represent the spectra using the same median values but with $c_1 = 0$. The energy density spectra derived from the free spectrum of the NANOGrav 15-year and EPTA DR2 new datasets are shown with blue and green shading.}
\end{figure}

As illustrated in Eq.~(\ref{eq:Php}), the energy density spectrum of second-order SIGWs is contingent on the second-order kernel function $I^{(2)}_{hf}$ and the primordial power spectrum $\mathcal{P}_{\zeta}(k)$. Here, the second-order kernel function $I^{(2)}_{hf}$  depends on the cosmological dominant era and the specific form of $f(R)$ theory. Thus, the energy density spectrum of second-order SIGWs is reliant on both the $f(R)$ theory 
 parameters and the primordial power spectrum. The constraints that \ac{PTA} data impose on the parameter space of the modified gravity theory are influenced by the form of the primordial power spectrum. In this paper, we focus on the HS model and the log-normal primordial power spectrum. The results in Eq.~(\ref{eq:Php}) are also valid for any form of $f(R)$ theory and primordial power spectrum. The energy density spectrum of SIGWs in the HS model is given in Fig.~\ref{fig:violin}. As shown in Fig.~\ref{fig:violin}, the parameter $c_1$ in the HS model affects the energy density spectrum of second order SIGWs. When the parameter $c_1=0$, our calculated spectrum matches the results obtained in general relativity. 

\subsection{PTA observation}
Using current PTA observational data, we analyze the parameter space of the HS model and the primordial power spectrum. Specifically, to constrain the parameter space of the primordial power spectrum and HS model in terms of PTA observations, we use Ceffyl \cite{lamb2023rapid} package embedded in PTArade \cite{mitridate2023ptarcade} to analyze the data from the first 14 frequency bins of NANOGrav 15-year dataset and the first 9 frequency bins of EPTA DR2 new dataset. 
\begin{figure}[htbp]
    \centering
    \includegraphics[width=.7\columnwidth]{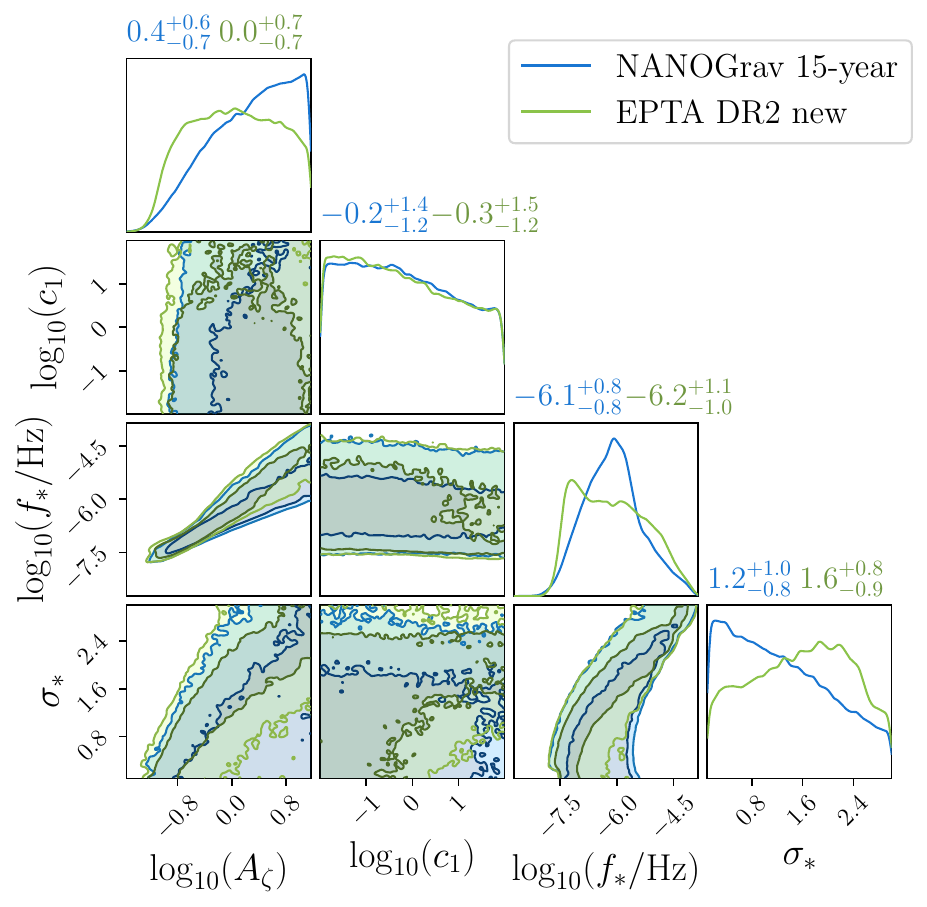}
\caption{\label{fig:corner} The corner plot of the posterior distributions. The blue and green solid curves correspond to the NANOGrav 15-year and EPTA DR2 new data sets, respectively. The contours in the off-diagonal panels denote the $1$-$\sigma$ and $2$-$\sigma$ ranges of the 2D posteriors. The numbers above the figures represent the median values and $1$-$\sigma$ ranges of the parameters.}
\end{figure}
We present the posteriors distributions in Fig.~\ref{fig:corner}, where the prior distributions of $\log_{10}(A_{\zeta})$, $\log_{10}(c_1)$, $\log_{10}(f_*/\mathrm{Hz})$, and $\sigma_*$ are set as uniform distributions over the intervals $[-3,1]$, $[-2,2]$, $[-10,-3]$, and $[0.1,3]$, respectively. Analyzing the posterior distributions of second order SIGWs energy density spectrum $\bar{\Omega}^{(2)}_{\mathrm{GW},0}(k)$, we find that for the parameters $A_{\zeta}$, $\log(f_*/\mathrm{Hz})$, and $\sigma_*$ in the primordial power spectrum, relatively accurate constraints can be provided. However, for the parameter $c_1$ in the modified gravity effects, the current \ac{PTA} observational data do not provide strong constraints. To precisely constrain the parameter space of the modified gravity model, we need to rely on more precise \ac{PTA} observational data and future joint constraints from \acp{GW} experiments in other frequency bands, such as \acp{LISA}.

\section{Conclusion and discussion}\label{sec:5}

In this paper, we systematically investigated the second order SIGWs in  $f(R)$ gravity, thoroughly discussing both first and second order cosmological perturbation equations in $f(R)$ theory. Our analysis reveals that, similar to the interaction between neutrinos and SIGWs, the presence of modified gravity effects generates additional anisotropic stress tensors as sources of \acp{GW}. Moreover, these effects modify the dynamic terms on the left-hand side of the equation of motion of second order SIGWs, resulting in a different form of the second order kernel functions compared to general relativity. By considering a log-normal primordial power spectrum, we calculated the energy density spectrum of the second order SIGWs in the HS model. Our results indicate that the effects of modified gravity has a substantial impact on the energy density spectrum of the second order SIGWs. 

Using current PTA observational data, we performed a detailed analysis of the parameter space for primordial power spectrum and HS model. The results of the Bayesian analysis indicate that the current \ac{PTA} observational data cannot exclude the presence of modified gravitational effects in the HS model, nor can they  effectively constrain its parameter space.  The theoretical framework presented in this paper is universally applicable to $f(R)$ modified gravity theories. The conclusions drawn from  Eq.~(\ref{eq:fs1})--Eq.~(\ref{eq:fs2}) and Eq.~(\ref{eq:ehfR})--Eq.~(\ref{eq:Ssigma}) are equally applicable to other types of $f(R)$ modified gravity models. By combining the theoretical results presented in this paper with more precise future PTA observational data, the parameter space of various $f(R)$ modified gravity models will be increasingly constrained in future research.

As previously discussed, the $f(R)$ theory introduces additional scalar modes, which in turn affect the polarization modes of \acp{GW} due to modified gravity effects. The Hellings–Downs curve from current PTA observations does not entirely exclude the presence of these extra modes \cite{NANOGrav:2023gor}. This paper focuses on the impact of modified gravity effects on the cosmological perturbation equations and the energy density spectrum of SIGWs. The conclusions related to second order SIGWs can be extended to other types of modified gravity theories \cite{Tzerefos:2023mpe,Zhang:2024vfw,Feng:2024yic,Papanikolaou:2022hkg,Domenech:2024drm}. Furthermore, in this paper, we neglected the potential impact of large primordial tensor perturbations and primordial non-Gaussianity on small scales, as well as the effects of higher order cosmological perturbations \cite{Chang:2023vjk}. Accounting for these effects would lead to substantial changes in the energy density spectrum of SIGWs. Related researches might be presented in the future.

\acknowledgments
The work is supported in part by the National Natural Science Foundation of China (NSFC) grants  No.12475075, No. 11935009, No. 12375052, No. 12075249, 11690022 and 12275276.

\bibliography{biblio}

\providecommand{\href}[2]{#2}\begingroup\raggedright\begin{thebibliography}{100}

\bibitem{LIGOScientific:2016aoc}
{\scshape LIGO Scientific, Virgo} collaboration, \emph{{Observation of Gravitational Waves from a Binary Black Hole Merger}}, \href{https://doi.org/10.1103/PhysRevLett.116.061102}{\emph{Phys. Rev. Lett.} {\bfseries 116} (2016) 061102} [\href{https://arxiv.org/abs/1602.03837}{{\ttfamily 1602.03837}}].

\bibitem{LIGOScientific:2016vlm}
{\scshape LIGO Scientific, Virgo} collaboration, \emph{{Properties of the Binary Black Hole Merger GW150914}}, \href{https://doi.org/10.1103/PhysRevLett.116.241102}{\emph{Phys. Rev. Lett.} {\bfseries 116} (2016) 241102} [\href{https://arxiv.org/abs/1602.03840}{{\ttfamily 1602.03840}}].

\bibitem{Annala:2017llu}
E.~Annala, T.~Gorda, A.~Kurkela and A.~Vuorinen, \emph{{Gravitational-wave constraints on the neutron-star-matter Equation of State}}, \href{https://doi.org/10.1103/PhysRevLett.120.172703}{\emph{Phys. Rev. Lett.} {\bfseries 120} (2018) 172703} [\href{https://arxiv.org/abs/1711.02644}{{\ttfamily 1711.02644}}].

\bibitem{LIGOScientific:2017adf}
{\scshape LIGO Scientific, Virgo, 1M2H, Dark Energy Camera GW-E, DES, DLT40, Las Cumbres Observatory, VINROUGE, MASTER} collaboration, \emph{{A gravitational-wave standard siren measurement of the Hubble constant}}, \href{https://doi.org/10.1038/nature24471}{\emph{Nature} {\bfseries 551} (2017) 85} [\href{https://arxiv.org/abs/1710.05835}{{\ttfamily 1710.05835}}].

\bibitem{BICEP:2021xfz}
{\scshape BICEP, Keck} collaboration, \emph{{Improved Constraints on Primordial Gravitational Waves using Planck, WMAP, and BICEP/Keck Observations through the 2018 Observing Season}}, \href{https://doi.org/10.1103/PhysRevLett.127.151301}{\emph{Phys. Rev. Lett.} {\bfseries 127} (2021) 151301} [\href{https://arxiv.org/abs/2110.00483}{{\ttfamily 2110.00483}}].

\bibitem{Ananda:2006af}
K.N.~Ananda, C.~Clarkson and D.~Wands, \emph{{The Cosmological gravitational wave background from primordial density perturbations}}, \href{https://doi.org/10.1103/PhysRevD.75.123518}{\emph{Phys. Rev. D} {\bfseries 75} (2007) 123518} [\href{https://arxiv.org/abs/gr-qc/0612013}{{\ttfamily gr-qc/0612013}}].

\bibitem{Domenech:2021ztg}
G.~Dom\`enech, \emph{{Scalar Induced Gravitational Waves Review}}, \href{https://doi.org/10.3390/universe7110398}{\emph{Universe} {\bfseries 7} (2021) 398} [\href{https://arxiv.org/abs/2109.01398}{{\ttfamily 2109.01398}}].

\bibitem{Kohri:2018awv}
K.~Kohri and T.~Terada, \emph{{Semianalytic calculation of gravitational wave spectrum nonlinearly induced from primordial curvature perturbations}}, \href{https://doi.org/10.1103/PhysRevD.97.123532}{\emph{Phys. Rev. D} {\bfseries 97} (2018) 123532} [\href{https://arxiv.org/abs/1804.08577}{{\ttfamily 1804.08577}}].

\bibitem{Mollerach:2003nq}
S.~Mollerach, D.~Harari and S.~Matarrese, \emph{{CMB polarization from secondary vector and tensor modes}}, \href{https://doi.org/10.1103/PhysRevD.69.063002}{\emph{Phys. Rev. D} {\bfseries 69} (2004) 063002} [\href{https://arxiv.org/abs/astro-ph/0310711}{{\ttfamily astro-ph/0310711}}].

\bibitem{NANOGrav:2023gor}
{\scshape NANOGrav} collaboration, \emph{{The NANOGrav 15 yr Data Set: Evidence for a Gravitational-wave Background}}, \href{https://doi.org/10.3847/2041-8213/acdac6}{\emph{Astrophys. J. Lett.} {\bfseries 951} (2023) L8} [\href{https://arxiv.org/abs/2306.16213}{{\ttfamily 2306.16213}}].

\bibitem{Reardon:2023gzh}
D.J.~Reardon et~al., \emph{{Search for an Isotropic Gravitational-wave Background with the Parkes Pulsar Timing Array}}, \href{https://doi.org/10.3847/2041-8213/acdd02}{\emph{Astrophys. J. Lett.} {\bfseries 951} (2023) L6} [\href{https://arxiv.org/abs/2306.16215}{{\ttfamily 2306.16215}}].

\bibitem{EPTA:2023fyk}
{\scshape EPTA} collaboration, \emph{{The second data release from the European Pulsar Timing Array III. Search for gravitational wave signals}}, \href{https://doi.org/10.1051/0004-6361/202346844}{\emph{Astron. Astrophys.} {\bfseries 678} (2023) A50} [\href{https://arxiv.org/abs/2306.16214}{{\ttfamily 2306.16214}}].

\bibitem{Xu:2023wog}
H.~Xu et~al., \emph{{Searching for the Nano-Hertz Stochastic Gravitational Wave Background with the Chinese Pulsar Timing Array Data Release I}}, \href{https://doi.org/10.1088/1674-4527/acdfa5}{\emph{Res. Astron. Astrophys.} {\bfseries 23} (2023) 075024} [\href{https://arxiv.org/abs/2306.16216}{{\ttfamily 2306.16216}}].

\bibitem{Wang:2019kaf}
S.~Wang, T.~Terada and K.~Kohri, \emph{{Prospective constraints on the primordial black hole abundance from the stochastic gravitational-wave backgrounds produced by coalescing events and curvature perturbations}}, \href{https://doi.org/10.1103/PhysRevD.99.103531}{\emph{Phys. Rev. D} {\bfseries 99} (2019) 103531} [\href{https://arxiv.org/abs/1903.05924}{{\ttfamily 1903.05924}}].

\bibitem{Byrnes:2018txb}
C.T.~Byrnes, P.S.~Cole and S.P.~Patil, \emph{{Steepest growth of the power spectrum and primordial black holes}}, \href{https://doi.org/10.1088/1475-7516/2019/06/028}{\emph{JCAP} {\bfseries 06} (2019) 028} [\href{https://arxiv.org/abs/1811.11158}{{\ttfamily 1811.11158}}].

\bibitem{Inomata:2020lmk}
K.~Inomata, M.~Kawasaki, K.~Mukaida, T.~Terada and T.T.~Yanagida, \emph{{Gravitational Wave Production right after a Primordial Black Hole Evaporation}}, \href{https://doi.org/10.1103/PhysRevD.101.123533}{\emph{Phys. Rev. D} {\bfseries 101} (2020) 123533} [\href{https://arxiv.org/abs/2003.10455}{{\ttfamily 2003.10455}}].

\bibitem{Ballesteros:2020qam}
G.~Ballesteros, J.~Rey, M.~Taoso and A.~Urbano, \emph{{Primordial black holes as dark matter and gravitational waves from single-field polynomial inflation}}, \href{https://doi.org/10.1088/1475-7516/2020/07/025}{\emph{JCAP} {\bfseries 07} (2020) 025} [\href{https://arxiv.org/abs/2001.08220}{{\ttfamily 2001.08220}}].

\bibitem{Lin:2020goi}
J.~Lin, Q.~Gao, Y.~Gong, Y.~Lu, C.~Zhang and F.~Zhang, \emph{{Primordial black holes and secondary gravitational waves from $k$ and $G$ inflation}}, \href{https://doi.org/10.1103/PhysRevD.101.103515}{\emph{Phys. Rev. D} {\bfseries 101} (2020) 103515} [\href{https://arxiv.org/abs/2001.05909}{{\ttfamily 2001.05909}}].

\bibitem{Chen:2019xse}
Z.-C.~Chen, C.~Yuan and Q.-G.~Huang, \emph{{Pulsar Timing Array Constraints on Primordial Black Holes with NANOGrav 11-Year Dataset}}, \href{https://doi.org/10.1103/PhysRevLett.124.251101}{\emph{Phys. Rev. Lett.} {\bfseries 124} (2020) 251101} [\href{https://arxiv.org/abs/1910.12239}{{\ttfamily 1910.12239}}].

\bibitem{Cai:2019elf}
R.-G.~Cai, S.~Pi, S.-J.~Wang and X.-Y.~Yang, \emph{{Pulsar Timing Array Constraints on the Induced Gravitational Waves}}, \href{https://doi.org/10.1088/1475-7516/2019/10/059}{\emph{JCAP} {\bfseries 10} (2019) 059} [\href{https://arxiv.org/abs/1907.06372}{{\ttfamily 1907.06372}}].

\bibitem{Cai:2019jah}
Y.-F.~Cai, C.~Chen, X.~Tong, D.-G.~Wang and S.-F.~Yan, \emph{{When Primordial Black Holes from Sound Speed Resonance Meet a Stochastic Background of Gravitational Waves}}, \href{https://doi.org/10.1103/PhysRevD.100.043518}{\emph{Phys. Rev. D} {\bfseries 100} (2019) 043518} [\href{https://arxiv.org/abs/1902.08187}{{\ttfamily 1902.08187}}].

\bibitem{Ando:2018qdb}
K.~Ando, K.~Inomata and M.~Kawasaki, \emph{{Primordial black holes and uncertainties in the choice of the window function}}, \href{https://doi.org/10.1103/PhysRevD.97.103528}{\emph{Phys. Rev. D} {\bfseries 97} (2018) 103528} [\href{https://arxiv.org/abs/1802.06393}{{\ttfamily 1802.06393}}].

\bibitem{Di:2017ndc}
H.~Di and Y.~Gong, \emph{{Primordial black holes and second order gravitational waves from ultra-slow-roll inflation}}, \href{https://doi.org/10.1088/1475-7516/2018/07/007}{\emph{JCAP} {\bfseries 07} (2018) 007} [\href{https://arxiv.org/abs/1707.09578}{{\ttfamily 1707.09578}}].

\bibitem{Gao:2021vxb}
Q.~Gao, \emph{{Primordial black holes and secondary gravitational waves from chaotic inflation}}, \href{https://doi.org/10.1007/s11433-021-1708-9}{\emph{Sci. China Phys. Mech. Astron.} {\bfseries 64} (2021) 280411} [\href{https://arxiv.org/abs/2102.07369}{{\ttfamily 2102.07369}}].

\bibitem{Changa:2022trj}
Z.~Chang, Y.-T.~Kuang, X.~Zhang and J.-Z.~Zhou, \emph{{Second order scalar perturbations induced by primordial curvature and tensor perturbations}},  \href{https://arxiv.org/abs/2211.11948}{{\ttfamily 2211.11948}}.

\bibitem{Zhou:2020kkf}
Z.~Zhou, J.~Jiang, Y.-F.~Cai, M.~Sasaki and S.~Pi, \emph{{Primordial black holes and gravitational waves from resonant amplification during inflation}}, \href{https://doi.org/10.1103/PhysRevD.102.103527}{\emph{Phys. Rev. D} {\bfseries 102} (2020) 103527} [\href{https://arxiv.org/abs/2010.03537}{{\ttfamily 2010.03537}}].

\bibitem{Cai:2021wzd}
R.-G.~Cai, C.~Chen and C.~Fu, \emph{{Primordial black holes and stochastic gravitational wave background from inflation with a noncanonical spectator field}}, \href{https://doi.org/10.1103/PhysRevD.104.083537}{\emph{Phys. Rev. D} {\bfseries 104} (2021) 083537} [\href{https://arxiv.org/abs/2108.03422}{{\ttfamily 2108.03422}}].

\bibitem{Hwang:2017oxa}
J.-C.~Hwang, D.~Jeong and H.~Noh, \emph{{Gauge dependence of gravitational waves generated from scalar perturbations}}, \href{https://doi.org/10.3847/1538-4357/aa74be}{\emph{Astrophys. J.} {\bfseries 842} (2017) 46} [\href{https://arxiv.org/abs/1704.03500}{{\ttfamily 1704.03500}}].

\bibitem{Yuan:2019fwv}
C.~Yuan, Z.-C.~Chen and Q.-G.~Huang, \emph{{Scalar induced gravitational waves in different gauges}}, \href{https://doi.org/10.1103/PhysRevD.101.063018}{\emph{Phys. Rev. D} {\bfseries 101} (2020) 063018} [\href{https://arxiv.org/abs/1912.00885}{{\ttfamily 1912.00885}}].

\bibitem{Inomata:2019yww}
K.~Inomata and T.~Terada, \emph{{Gauge Independence of Induced Gravitational Waves}}, \href{https://doi.org/10.1103/PhysRevD.101.023523}{\emph{Phys. Rev. D} {\bfseries 101} (2020) 023523} [\href{https://arxiv.org/abs/1912.00785}{{\ttfamily 1912.00785}}].

\bibitem{DeLuca:2019ufz}
V.~De~Luca, G.~Franciolini, A.~Kehagias and A.~Riotto, \emph{{On the Gauge Invariance of Cosmological Gravitational Waves}}, \href{https://doi.org/10.1088/1475-7516/2020/03/014}{\emph{JCAP} {\bfseries 03} (2020) 014} [\href{https://arxiv.org/abs/1911.09689}{{\ttfamily 1911.09689}}].

\bibitem{Domenech:2020xin}
G.~Dom\`enech and M.~Sasaki, \emph{{Approximate gauge independence of the induced gravitational wave spectrum}}, \href{https://doi.org/10.1103/PhysRevD.103.063531}{\emph{Phys. Rev. D} {\bfseries 103} (2021) 063531} [\href{https://arxiv.org/abs/2012.14016}{{\ttfamily 2012.14016}}].

\bibitem{Chang:2020tji}
Z.~Chang, S.~Wang and Q.-H.~Zhu, \emph{{Note on gauge invariance of second order cosmological perturbations}}, \href{https://doi.org/10.1088/1674-1137/ac0c74}{\emph{Chin. Phys. C} {\bfseries 45} (2021) 095101} [\href{https://arxiv.org/abs/2009.11025}{{\ttfamily 2009.11025}}].

\bibitem{Ali:2020sfw}
A.~Ali, Y.~Gong and Y.~Lu, \emph{{Gauge transformation of scalar induced tensor perturbation during matter domination}}, \href{https://doi.org/10.1103/PhysRevD.103.043516}{\emph{Phys. Rev. D} {\bfseries 103} (2021) 043516} [\href{https://arxiv.org/abs/2009.11081}{{\ttfamily 2009.11081}}].

\bibitem{Lu:2020diy}
Y.~Lu, A.~Ali, Y.~Gong, J.~Lin and F.~Zhang, \emph{{Gauge transformation of scalar induced gravitational waves}}, \href{https://doi.org/10.1103/PhysRevD.102.083503(2020)}{\emph{Phys. Rev. D} {\bfseries 102} (2020) 083503} [\href{https://arxiv.org/abs/2006.03450}{{\ttfamily 2006.03450}}].

\bibitem{Tomikawa:2019tvi}
K.~Tomikawa and T.~Kobayashi, \emph{{Gauge dependence of gravitational waves generated at second order from scalar perturbations}}, \href{https://doi.org/10.1103/PhysRevD.101.083529}{\emph{Phys. Rev. D} {\bfseries 101} (2020) 083529} [\href{https://arxiv.org/abs/1910.01880}{{\ttfamily 1910.01880}}].

\bibitem{Gurian:2021rfv}
J.~Gurian, D.~Jeong, J.-c.~Hwang and H.~Noh, \emph{{Gauge-invariant tensor perturbations induced from baryon-CDM relative velocity and the B-mode polarization of the CMB}}, \href{https://doi.org/10.1103/PhysRevD.104.083534}{\emph{Phys. Rev. D} {\bfseries 104} (2021) 083534} [\href{https://arxiv.org/abs/2104.03330}{{\ttfamily 2104.03330}}].

\bibitem{Uggla:2018fiy}
C.~Uggla and J.~Wainwright, \emph{{Second order cosmological perturbations: simplified gauge change formulas}}, \href{https://doi.org/10.1088/1361-6382/aaf924}{\emph{Class. Quant. Grav.} {\bfseries 36} (2019) 035004} [\href{https://arxiv.org/abs/1801.04300}{{\ttfamily 1801.04300}}].

\bibitem{Ali:2023moi}
A.~Ali, Y.-P.~Hu, M.~Sabir and T.~Sui, \emph{{On the gauge dependence of scalar induced secondary gravitational waves during radiation and matter domination eras}}, \href{https://doi.org/10.1007/s11433-022-2118-5}{\emph{Sci. China Phys. Mech. Astron.} {\bfseries 66} (2023) 290411} [\href{https://arxiv.org/abs/2308.04713}{{\ttfamily 2308.04713}}].

\bibitem{Papanikolaou:2020qtd}
T.~Papanikolaou, V.~Vennin and D.~Langlois, \emph{{Gravitational waves from a universe filled with primordial black holes}}, \href{https://doi.org/10.1088/1475-7516/2021/03/053}{\emph{JCAP} {\bfseries 03} (2021) 053} [\href{https://arxiv.org/abs/2010.11573}{{\ttfamily 2010.11573}}].

\bibitem{Domenech:2020kqm}
G.~Dom\`enech, S.~Pi and M.~Sasaki, \emph{{Induced gravitational waves as a probe of thermal history of the universe}}, \href{https://doi.org/10.1088/1475-7516/2020/08/017}{\emph{JCAP} {\bfseries 08} (2020) 017} [\href{https://arxiv.org/abs/2005.12314}{{\ttfamily 2005.12314}}].

\bibitem{Domenech:2019quo}
G.~Dom\`enech, \emph{{Induced gravitational waves in a general cosmological background}}, \href{https://doi.org/10.1142/S0218271820500285}{\emph{Int. J. Mod. Phys. D} {\bfseries 29} (2020) 2050028} [\href{https://arxiv.org/abs/1912.05583}{{\ttfamily 1912.05583}}].

\bibitem{Inomata:2019zqy}
K.~Inomata, K.~Kohri, T.~Nakama and T.~Terada, \emph{{Gravitational Waves Induced by Scalar Perturbations during a Gradual Transition from an Early Matter Era to the Radiation Era}}, \href{https://doi.org/10.1088/1475-7516/2019/10/071}{\emph{JCAP} {\bfseries 10} (2019) 071} [\href{https://arxiv.org/abs/1904.12878}{{\ttfamily 1904.12878}}].

\bibitem{Inomata:2019ivs}
K.~Inomata, K.~Kohri, T.~Nakama and T.~Terada, \emph{{Enhancement of Gravitational Waves Induced by Scalar Perturbations due to a Sudden Transition from an Early Matter Era to the Radiation Era}}, \href{https://doi.org/10.1103/PhysRevD.108.049901}{\emph{Phys. Rev. D} {\bfseries 100} (2019) 043532} [\href{https://arxiv.org/abs/1904.12879}{{\ttfamily 1904.12879}}].

\bibitem{Witkowski:2021raz}
L.T.~Witkowski, G.~Dom\`enech, J.~Fumagalli and S.~Renaux-Petel, \emph{{Expansion history-dependent oscillations in the scalar-induced gravitational wave background}}, \href{https://doi.org/10.1088/1475-7516/2022/05/028}{\emph{JCAP} {\bfseries 05} (2022) 028} [\href{https://arxiv.org/abs/2110.09480}{{\ttfamily 2110.09480}}].

\bibitem{Dalianis:2020gup}
I.~Dalianis and C.~Kouvaris, \emph{{Gravitational waves from density perturbations in an early matter domination era}}, \href{https://doi.org/10.1088/1475-7516/2021/07/046}{\emph{JCAP} {\bfseries 07} (2021) 046} [\href{https://arxiv.org/abs/2012.09255}{{\ttfamily 2012.09255}}].

\bibitem{Hajkarim:2019nbx}
F.~Hajkarim and J.~Schaffner-Bielich, \emph{{Thermal History of the Early Universe and Primordial Gravitational Waves from Induced Scalar Perturbations}}, \href{https://doi.org/10.1103/PhysRevD.101.043522}{\emph{Phys. Rev. D} {\bfseries 101} (2020) 043522} [\href{https://arxiv.org/abs/1910.12357}{{\ttfamily 1910.12357}}].

\bibitem{Bernal:2019lpc}
N.~Bernal and F.~Hajkarim, \emph{{Primordial Gravitational Waves in Nonstandard Cosmologies}}, \href{https://doi.org/10.1103/PhysRevD.100.063502}{\emph{Phys. Rev. D} {\bfseries 100} (2019) 063502} [\href{https://arxiv.org/abs/1905.10410}{{\ttfamily 1905.10410}}].

\bibitem{Das:2021wad}
S.~Das, A.~Maharana and F.~Muia, \emph{{A faster growth of perturbations in an early matter dominated epoch: primordial black holes and gravitational waves}}, \href{https://doi.org/10.1093/mnras/stac1620}{\emph{Mon. Not. Roy. Astron. Soc.} {\bfseries 515} (2022) 13} [\href{https://arxiv.org/abs/2112.11486}{{\ttfamily 2112.11486}}].

\bibitem{Haque:2021dha}
M.R.~Haque, D.~Maity, T.~Paul and L.~Sriramkumar, \emph{{Decoding the phases of early and late time reheating through imprints on primordial gravitational waves}}, \href{https://doi.org/10.1103/PhysRevD.104.063513}{\emph{Phys. Rev. D} {\bfseries 104} (2021) 063513} [\href{https://arxiv.org/abs/2105.09242}{{\ttfamily 2105.09242}}].

\bibitem{Domenech:2020ssp}
G.~Dom\`enech, C.~Lin and M.~Sasaki, \emph{{Gravitational wave constraints on the primordial black hole dominated early universe}}, \href{https://doi.org/10.1088/1475-7516/2021/11/E01}{\emph{JCAP} {\bfseries 04} (2021) 062} [\href{https://arxiv.org/abs/2012.08151}{{\ttfamily 2012.08151}}].

\bibitem{Domenech:2021and}
G.~Dom\`enech, S.~Passaglia and S.~Renaux-Petel, \emph{{Gravitational waves from dark matter isocurvature}}, \href{https://doi.org/10.1088/1475-7516/2022/03/023}{\emph{JCAP} {\bfseries 03} (2022) 023} [\href{https://arxiv.org/abs/2112.10163}{{\ttfamily 2112.10163}}].

\bibitem{Liu:2023pau}
L.~Liu, Z.-C.~Chen and Q.-G.~Huang, \emph{{Probing the equation of state of the early Universe with pulsar timing arrays}},  \href{https://arxiv.org/abs/2307.14911}{{\ttfamily 2307.14911}}.

\bibitem{Mangilli:2008bw}
A.~Mangilli, N.~Bartolo, S.~Matarrese and A.~Riotto, \emph{{The impact of cosmic neutrinos on the gravitational-wave background}}, \href{https://doi.org/10.1103/PhysRevD.78.083517}{\emph{Phys. Rev. D} {\bfseries 78} (2008) 083517} [\href{https://arxiv.org/abs/0805.3234}{{\ttfamily 0805.3234}}].

\bibitem{Saga:2014jca}
S.~Saga, K.~Ichiki and N.~Sugiyama, \emph{{Impact of anisotropic stress of free-streaming particles on gravitational waves induced by cosmological density perturbations}}, \href{https://doi.org/10.1103/PhysRevD.91.024030}{\emph{Phys. Rev. D} {\bfseries 91} (2015) 024030} [\href{https://arxiv.org/abs/1412.1081}{{\ttfamily 1412.1081}}].

\bibitem{Zhang:2022dgx}
X.~Zhang, J.-Z.~Zhou and Z.~Chang, \emph{{Impact of the free-streaming neutrinos to the second order induced gravitational waves}}, \href{https://doi.org/10.1140/epjc/s10052-022-10742-x}{\emph{Eur. Phys. J. C} {\bfseries 82} (2022) 781} [\href{https://arxiv.org/abs/2208.12948}{{\ttfamily 2208.12948}}].

\bibitem{Yuan:2023ofl}
C.~Yuan, D.-S.~Meng and Q.-G.~Huang, \emph{{Full analysis of the scalar-induced gravitational waves for the curvature perturbation with local-type non-Gaussianities}},  \href{https://arxiv.org/abs/2308.07155}{{\ttfamily 2308.07155}}.

\bibitem{Yu:2024xmz}
Y.-H.~Yu and S.~Wang, \emph{{Silk Damping in Scalar-Induced Gravitational Waves: A Novel Probe for New Physics}},  \href{https://arxiv.org/abs/2405.02960}{{\ttfamily 2405.02960}}.

\bibitem{Cai:2018dig}
R.-g.~Cai, S.~Pi and M.~Sasaki, \emph{{Gravitational Waves Induced by non-Gaussian Scalar Perturbations}}, \href{https://doi.org/10.1103/PhysRevLett.122.201101}{\emph{Phys. Rev. Lett.} {\bfseries 122} (2019) 201101} [\href{https://arxiv.org/abs/1810.11000}{{\ttfamily 1810.11000}}].

\bibitem{Atal:2021jyo}
V.~Atal and G.~Dom\`enech, \emph{{Probing non-Gaussianities with the high frequency tail of induced gravitational waves}}, \href{https://doi.org/10.1088/1475-7516/2021/06/001}{\emph{JCAP} {\bfseries 06} (2021) 001} [\href{https://arxiv.org/abs/2103.01056}{{\ttfamily 2103.01056}}].

\bibitem{Zhang:2020uek}
F.~Zhang, Y.~Gong, J.~Lin, Y.~Lu and Z.~Yi, \emph{{Primordial non-Gaussianity from G-inflation}}, \href{https://doi.org/10.1088/1475-7516/2021/04/045}{\emph{JCAP} {\bfseries 04} (2021) 045} [\href{https://arxiv.org/abs/2012.06960}{{\ttfamily 2012.06960}}].

\bibitem{Yuan:2020iwf}
C.~Yuan and Q.-G.~Huang, \emph{{Gravitational waves induced by the local-type non-Gaussian curvature perturbations}}, \href{https://doi.org/10.1016/j.physletb.2021.136606}{\emph{Phys. Lett. B} {\bfseries 821} (2021) 136606} [\href{https://arxiv.org/abs/2007.10686}{{\ttfamily 2007.10686}}].

\bibitem{Davies:2021loj}
M.W.~Davies, P.~Carrilho and D.J.~Mulryne, \emph{{Non-Gaussianity in inflationary scenarios for primordial black holes}}, \href{https://doi.org/10.1088/1475-7516/2022/06/019}{\emph{JCAP} {\bfseries 06} (2022) 019} [\href{https://arxiv.org/abs/2110.08189}{{\ttfamily 2110.08189}}].

\bibitem{Rezazadeh:2021clf}
K.~Rezazadeh, Z.~Teimoori, S.~Karimi and K.~Karami, \emph{{Non-Gaussianity and secondary gravitational waves from primordial black holes production in $\alpha $-attractor inflation}}, \href{https://doi.org/10.1140/epjc/s10052-022-10735-w}{\emph{Eur. Phys. J. C} {\bfseries 82} (2022) 758} [\href{https://arxiv.org/abs/2110.01482}{{\ttfamily 2110.01482}}].

\bibitem{Kristiano:2021urj}
J.~Kristiano and J.~Yokoyama, \emph{{Why Must Primordial Non-Gaussianity Be Very Small?}}, \href{https://doi.org/10.1103/PhysRevLett.128.061301}{\emph{Phys. Rev. Lett.} {\bfseries 128} (2022) 061301} [\href{https://arxiv.org/abs/2104.01953}{{\ttfamily 2104.01953}}].

\bibitem{Bartolo:2018qqn}
N.~Bartolo, V.~Domcke, D.G.~Figueroa, J.~Garc\'\i{}a-Bellido, M.~Peloso, M.~Pieroni et~al., \emph{{Probing non-Gaussian Stochastic Gravitational Wave Backgrounds with LISA}}, \href{https://doi.org/10.1088/1475-7516/2018/11/034}{\emph{JCAP} {\bfseries 11} (2018) 034} [\href{https://arxiv.org/abs/1806.02819}{{\ttfamily 1806.02819}}].

\bibitem{Adshead:2021hnm}
P.~Adshead, K.D.~Lozanov and Z.J.~Weiner, \emph{{Non-Gaussianity and the induced gravitational wave background}}, \href{https://doi.org/10.1088/1475-7516/2021/10/080}{\emph{JCAP} {\bfseries 10} (2021) 080} [\href{https://arxiv.org/abs/2105.01659}{{\ttfamily 2105.01659}}].

\bibitem{Li:2023qua}
J.-P.~Li, S.~Wang, Z.-C.~Zhao and K.~Kohri, \emph{{Primordial non-Gaussianity f $_{NL}$ and anisotropies in scalar-induced gravitational waves}}, \href{https://doi.org/10.1088/1475-7516/2023/10/056}{\emph{JCAP} {\bfseries 10} (2023) 056} [\href{https://arxiv.org/abs/2305.19950}{{\ttfamily 2305.19950}}].

\bibitem{Li:2023xtl}
J.-P.~Li, S.~Wang, Z.-C.~Zhao and K.~Kohri, \emph{{Complete Analysis of Scalar-Induced Gravitational Waves and Primordial Non-Gaussianities $f_{\mathrm{NL}}$ and $g_{\mathrm{NL}}$}},  \href{https://arxiv.org/abs/2309.07792}{{\ttfamily 2309.07792}}.

\bibitem{Garcia-Saenz:2022tzu}
S.~Garcia-Saenz, L.~Pinol, S.~Renaux-Petel and D.~Werth, \emph{{No-go theorem for scalar-trispectrum-induced gravitational waves}}, \href{https://doi.org/10.1088/1475-7516/2023/03/057}{\emph{JCAP} {\bfseries 03} (2023) 057} [\href{https://arxiv.org/abs/2207.14267}{{\ttfamily 2207.14267}}].

\bibitem{Zhou:2021vcw}
J.-Z.~Zhou, X.~Zhang, Q.-H.~Zhu and Z.~Chang, \emph{{The third order scalar induced gravitational waves}}, \href{https://doi.org/10.1088/1475-7516/2022/05/013}{\emph{JCAP} {\bfseries 05} (2022) 013} [\href{https://arxiv.org/abs/2106.01641}{{\ttfamily 2106.01641}}].

\bibitem{Chang:2023vjk}
Z.~Chang, Y.-T.~Kuang, D.~Wu and J.-Z.~Zhou, \emph{{Probing scalar induced gravitational waves with PTA and LISA: the importance of third order correction}}, \href{https://doi.org/10.1088/1475-7516/2024/04/044}{\emph{JCAP} {\bfseries 2024} (2024) 044} [\href{https://arxiv.org/abs/2312.14409}{{\ttfamily 2312.14409}}].

\bibitem{Wang:2023sij}
S.~Wang, Z.-C.~Zhao and Q.-H.~Zhu, \emph{{Constraints on scalar-induced gravitational waves up to third order from a joint analysis of BBN, CMB, and PTA data}}, \href{https://doi.org/10.1103/PhysRevResearch.6.013207}{\emph{Phys. Rev. Res.} {\bfseries 6} (2024) 013207} [\href{https://arxiv.org/abs/2307.03095}{{\ttfamily 2307.03095}}].

\bibitem{Chang:2022nzu}
Z.~Chang, Y.-T.~Kuang, X.~Zhang and J.-Z.~Zhou, \emph{{Primordial black holes and third order scalar induced gravitational waves*}}, \href{https://doi.org/10.1088/1674-1137/acc649}{\emph{Chin. Phys. C} {\bfseries 47} (2023) 055104} [\href{https://arxiv.org/abs/2209.12404}{{\ttfamily 2209.12404}}].

\bibitem{Chang:2022dhh}
Z.~Chang, X.~Zhang and J.-Z.~Zhou, \emph{{The cosmological vector modes from a monochromatic primordial power spectrum}}, \href{https://doi.org/10.1088/1475-7516/2022/10/084}{\emph{JCAP} {\bfseries 10} (2022) 084} [\href{https://arxiv.org/abs/2207.01231}{{\ttfamily 2207.01231}}].

\bibitem{Zhou:2024ncc}
J.-Z.~Zhou, Y.-T.~Kuang, D.~Wu, H.~L\"u and Z.~Chang, \emph{{Induced gravitational waves for arbitrary higher orders: vertex rules and loop diagrams in cosmological perturbation theory}},  \href{https://arxiv.org/abs/2408.14052}{{\ttfamily 2408.14052}}.

\bibitem{Koyama:2015vza}
K.~Koyama, \emph{{Cosmological Tests of Modified Gravity}}, \href{https://doi.org/10.1088/0034-4885/79/4/046902}{\emph{Rept. Prog. Phys.} {\bfseries 79} (2016) 046902} [\href{https://arxiv.org/abs/1504.04623}{{\ttfamily 1504.04623}}].

\bibitem{Arjona:2018jhh}
R.~Arjona, W.~Cardona and S.~Nesseris, \emph{{Unraveling the effective fluid approach for $f(R)$ models in the subhorizon approximation}}, \href{https://doi.org/10.1103/PhysRevD.99.043516}{\emph{Phys. Rev. D} {\bfseries 99} (2019) 043516} [\href{https://arxiv.org/abs/1811.02469}{{\ttfamily 1811.02469}}].

\bibitem{Hu:2007nk}
W.~Hu and I.~Sawicki, \emph{{Models of f(R) Cosmic Acceleration that Evade Solar-System Tests}}, \href{https://doi.org/10.1103/PhysRevD.76.064004}{\emph{Phys. Rev. D} {\bfseries 76} (2007) 064004} [\href{https://arxiv.org/abs/0705.1158}{{\ttfamily 0705.1158}}].

\bibitem{Graham:2015rva}
P.W.~Graham, J.~Mardon and S.~Rajendran, \emph{{Vector Dark Matter from Inflationary Fluctuations}}, \href{https://doi.org/10.1103/PhysRevD.93.103520}{\emph{Phys. Rev. D} {\bfseries 93} (2016) 103520} [\href{https://arxiv.org/abs/1504.02102}{{\ttfamily 1504.02102}}].

\bibitem{Okano:2020uyr}
S.~Okano and T.~Fujita, \emph{{Chiral Gravitational Waves Produced in a Helical Magnetogenesis Model}}, \href{https://doi.org/10.1088/1475-7516/2021/03/026}{\emph{JCAP} {\bfseries 03} (2021) 026} [\href{https://arxiv.org/abs/2005.13833}{{\ttfamily 2005.13833}}].

\bibitem{Cai:2020ovp}
Y.-F.~Cai, C.~Lin, B.~Wang and S.-F.~Yan, \emph{{Sound speed resonance of the stochastic gravitational wave background}}, \href{https://doi.org/10.1103/PhysRevLett.126.071303}{\emph{Phys. Rev. Lett.} {\bfseries 126} (2021) 071303} [\href{https://arxiv.org/abs/2009.09833}{{\ttfamily 2009.09833}}].

\bibitem{Gorji:2023ziy}
M.A.~Gorji and M.~Sasaki, \emph{{Primordial-tensor-induced stochastic gravitational waves}}, \href{https://doi.org/10.1016/j.physletb.2023.138236}{\emph{Phys. Lett. B} {\bfseries 846} (2023) 138236} [\href{https://arxiv.org/abs/2302.14080}{{\ttfamily 2302.14080}}].

\bibitem{Chang:2022vlv}
Z.~Chang, X.~Zhang and J.-Z.~Zhou, \emph{{Gravitational waves from primordial scalar and tensor perturbations}}, \href{https://doi.org/10.1103/PhysRevD.107.063510}{\emph{Phys. Rev. D} {\bfseries 107} (2023) 063510} [\href{https://arxiv.org/abs/2209.07693}{{\ttfamily 2209.07693}}].

\bibitem{Inomata:2020cck}
K.~Inomata, \emph{{Analytic solutions of scalar perturbations induced by scalar perturbations}}, \href{https://doi.org/10.1088/1475-7516/2021/03/013}{\emph{JCAP} {\bfseries 03} (2021) 013} [\href{https://arxiv.org/abs/2008.12300}{{\ttfamily 2008.12300}}].

\bibitem{OMurchadha:1973byk}
N.~O'Murchadha and J.W.~York, Jr., \emph{{Existence and uniqueness of solutions of the hamiltonian constraint on compact manifolds}}, \href{https://doi.org/10.1063/1.1666225}{\emph{J. Math. Phys.} {\bfseries 14} (1973) 1551}.

\bibitem{Weinberg:2003ur}
S.~Weinberg, \emph{{Damping of tensor modes in cosmology}}, \href{https://doi.org/10.1103/PhysRevD.69.023503}{\emph{Phys. Rev. D} {\bfseries 69} (2004) 023503} [\href{https://arxiv.org/abs/astro-ph/0306304}{{\ttfamily astro-ph/0306304}}].

\bibitem{Tsujikawa:2007gd}
S.~Tsujikawa, \emph{{Matter density perturbations and effective gravitational constant in modified gravity models of dark energy}}, \href{https://doi.org/10.1103/PhysRevD.76.023514}{\emph{Phys. Rev. D} {\bfseries 76} (2007) 023514} [\href{https://arxiv.org/abs/0705.1032}{{\ttfamily 0705.1032}}].

\bibitem{Papanikolaou:2021uhe}
T.~Papanikolaou, C.~Tzerefos, S.~Basilakos and E.N.~Saridakis, \emph{{Scalar induced gravitational waves from primordial black hole Poisson fluctuations in f(R) gravity}}, \href{https://doi.org/10.1088/1475-7516/2022/10/013}{\emph{JCAP} {\bfseries 10} (2022) 013} [\href{https://arxiv.org/abs/2112.15059}{{\ttfamily 2112.15059}}].

\bibitem{Olmo:2011uz}
G.J.~Olmo, \emph{{Palatini Approach to Modified Gravity: f(R) Theories and Beyond}}, \href{https://doi.org/10.1142/S0218271811018925}{\emph{Int. J. Mod. Phys. D} {\bfseries 20} (2011) 413} [\href{https://arxiv.org/abs/1101.3864}{{\ttfamily 1101.3864}}].

\bibitem{DeFelice:2010aj}
A.~De~Felice and S.~Tsujikawa, \emph{{f(R) theories}}, \href{https://doi.org/10.12942/lrr-2010-3}{\emph{Living Rev. Rel.} {\bfseries 13} (2010) 3} [\href{https://arxiv.org/abs/1002.4928}{{\ttfamily 1002.4928}}].

\bibitem{Arjona:2019rfn}
R.~Arjona, W.~Cardona and S.~Nesseris, \emph{{Designing Horndeski and the effective fluid approach}}, \href{https://doi.org/10.1103/PhysRevD.100.063526}{\emph{Phys. Rev. D} {\bfseries 100} (2019) 063526} [\href{https://arxiv.org/abs/1904.06294}{{\ttfamily 1904.06294}}].

\bibitem{Arjona:2020gtm}
R.~Arjona, \emph{{The effective fluid approach for modified gravity}},  in \emph{{2nd Hermann Minkowski Meeting on the Foundations of Spacetime Physics}}, 10, 2020 [\href{https://arxiv.org/abs/2010.04764}{{\ttfamily 2010.04764}}].

\bibitem{Cardona:2022lcz}
W.~Cardona, J.B.~Orjuela-Quintana and C.A.~Valenzuela-Toledo, \emph{{An effective fluid description of scalar-vector-tensor theories under the sub-horizon and quasi-static approximations}}, \href{https://doi.org/10.1088/1475-7516/2022/08/059}{\emph{JCAP} {\bfseries 08} (2022) 059} [\href{https://arxiv.org/abs/2206.02895}{{\ttfamily 2206.02895}}].

\bibitem{Starobinsky:1980te}
A.A.~Starobinsky, \emph{{A New Type of Isotropic Cosmological Models Without Singularity}}, \href{https://doi.org/10.1016/0370-2693(80)90670-X}{\emph{Phys. Lett. B} {\bfseries 91} (1980) 99}.

\bibitem{Katsuragawa:2019uto}
T.~Katsuragawa, T.~Nakamura, T.~Ikeda and S.~Capozziello, \emph{{Gravitational Waves in $F(R)$ Gravity: Scalar Waves and the Chameleon Mechanism}}, \href{https://doi.org/10.1103/PhysRevD.99.124050}{\emph{Phys. Rev. D} {\bfseries 99} (2019) 124050} [\href{https://arxiv.org/abs/1902.02494}{{\ttfamily 1902.02494}}].

\bibitem{Myung:2016zdl}
Y.S.~Myung, \emph{{Propagating Degrees of Freedom in $f(R)$ Gravity}}, \href{https://doi.org/10.1155/2016/3901734}{\emph{Adv. High Energy Phys.} {\bfseries 2016} (2016) 3901734} [\href{https://arxiv.org/abs/1608.01764}{{\ttfamily 1608.01764}}].

\bibitem{Gong:2017bru}
Y.~Gong and S.~Hou, \emph{{Gravitational Wave Polarizations in $f(R)$ Gravity and Scalar-Tensor Theory}}, \href{https://doi.org/10.1051/epjconf/201816801003}{\emph{EPJ Web Conf.} {\bfseries 168} (2018) 01003} [\href{https://arxiv.org/abs/1709.03313}{{\ttfamily 1709.03313}}].

\bibitem{Moretti:2019yhs}
F.~Moretti, F.~Bombacigno and G.~Montani, \emph{{Gauge invariant formulation of metric $f(R)$ gravity for gravitational waves}}, \href{https://doi.org/10.1103/PhysRevD.100.084014}{\emph{Phys. Rev. D} {\bfseries 100} (2019) 084014} [\href{https://arxiv.org/abs/1906.01899}{{\ttfamily 1906.01899}}].

\bibitem{Vainio:2016qas}
J.~Vainio and I.~Vilja, \emph{{$f(R)$ gravity constraints from gravitational waves}}, \href{https://doi.org/10.1007/s10714-017-2262-3}{\emph{Gen. Rel. Grav.} {\bfseries 49} (2017) 99} [\href{https://arxiv.org/abs/1603.09551}{{\ttfamily 1603.09551}}].

\bibitem{Saikawa:2018rcs}
K.~Saikawa and S.~Shirai, \emph{{Primordial gravitational waves, precisely: The role of thermodynamics in the Standard Model}}, \href{https://doi.org/10.1088/1475-7516/2018/05/035}{\emph{JCAP} {\bfseries 05} (2018) 035} [\href{https://arxiv.org/abs/1803.01038}{{\ttfamily 1803.01038}}].

\bibitem{Planck:2018vyg}
{\scshape Planck} collaboration, \emph{{Planck 2018 results. VI. Cosmological parameters}}, \href{https://doi.org/10.1051/0004-6361/201833910}{\emph{Astron. Astrophys.} {\bfseries 641} (2020) A6} [\href{https://arxiv.org/abs/1807.06209}{{\ttfamily 1807.06209}}].

\bibitem{Assadullahi:2009nf}
H.~Assadullahi and D.~Wands, \emph{{Gravitational waves from an early matter era}}, \href{https://doi.org/10.1103/PhysRevD.79.083511}{\emph{Phys. Rev. D} {\bfseries 79} (2009) 083511} [\href{https://arxiv.org/abs/0901.0989}{{\ttfamily 0901.0989}}].

\bibitem{Alabidi:2013lya}
L.~Alabidi, K.~Kohri, M.~Sasaki and Y.~Sendouda, \emph{{Observable induced gravitational waves from an early matter phase}}, \href{https://doi.org/10.1088/1475-7516/2013/05/033}{\emph{JCAP} {\bfseries 05} (2013) 033} [\href{https://arxiv.org/abs/1303.4519}{{\ttfamily 1303.4519}}].

\bibitem{Papanikolaou:2024kjb}
T.~Papanikolaou, X.-C.~He, X.-H.~Ma, Y.-F.~Cai, E.N.~Saridakis and M.~Sasaki, \emph{{New probe of non-Gaussianities with primordial black hole induced gravitational waves}}, \href{https://doi.org/10.1016/j.physletb.2024.138997}{\emph{Phys. Lett. B} {\bfseries 857} (2024) 138997} [\href{https://arxiv.org/abs/2403.00660}{{\ttfamily 2403.00660}}].

\bibitem{Papanikolaou:2022chm}
T.~Papanikolaou, \emph{{Gravitational waves induced from primordial black hole fluctuations: the~effect of an extended mass function}}, \href{https://doi.org/10.1088/1475-7516/2022/10/089}{\emph{JCAP} {\bfseries 10} (2022) 089} [\href{https://arxiv.org/abs/2207.11041}{{\ttfamily 2207.11041}}].

\bibitem{lamb2023rapid}
W.G.~Lamb, S.R.~Taylor and R.~van Haasteren, \emph{Rapid refitting techniques for bayesian spectral characterization of the gravitational wave background using pulsar timing arrays}, {\emph{Physical Review D} {\bfseries 108} (2023) 103019}.

\bibitem{mitridate2023ptarcade}
A.~Mitridate, D.~Wright, R.~von Eckardstein, T.~Schröder, J.~Nay, K.~Olum et~al., \emph{Ptarcade},  2023.

\bibitem{Tzerefos:2023mpe}
C.~Tzerefos, T.~Papanikolaou, E.N.~Saridakis and S.~Basilakos, \emph{{Scalar induced gravitational waves in modified teleparallel gravity theories}}, \href{https://doi.org/10.1103/PhysRevD.107.124019}{\emph{Phys. Rev. D} {\bfseries 107} (2023) 124019} [\href{https://arxiv.org/abs/2303.16695}{{\ttfamily 2303.16695}}].

\bibitem{Zhang:2024vfw}
F.~Zhang, J.-X.~Feng and X.~Gao, \emph{{Scalar induced gravitational waves in metric teleparallel gravity with the Nieh-Yan term}}, \href{https://doi.org/10.1103/PhysRevD.110.023537}{\emph{Phys. Rev. D} {\bfseries 110} (2024) 023537} [\href{https://arxiv.org/abs/2404.02922}{{\ttfamily 2404.02922}}].

\bibitem{Feng:2024yic}
J.-X.~Feng, F.~Zhang and X.~Gao, \emph{{Scalar induced gravitational waves in chiral scalar\textendash{}tensor theory of gravity}}, \href{https://doi.org/10.1140/epjc/s10052-024-13097-7}{\emph{Eur. Phys. J. C} {\bfseries 84} (2024) 736} [\href{https://arxiv.org/abs/2404.05289}{{\ttfamily 2404.05289}}].

\bibitem{Papanikolaou:2022hkg}
T.~Papanikolaou, C.~Tzerefos, S.~Basilakos and E.N.~Saridakis, \emph{{No constraints for f(T) gravity from gravitational waves induced from primordial black hole fluctuations}}, \href{https://doi.org/10.1140/epjc/s10052-022-11157-4}{\emph{Eur. Phys. J. C} {\bfseries 83} (2023) 31} [\href{https://arxiv.org/abs/2205.06094}{{\ttfamily 2205.06094}}].

\bibitem{Domenech:2024drm}
G.~Dom\`enech and A.~Ganz, \emph{{Enhanced induced gravitational waves in Horndeski gravity}},  \href{https://arxiv.org/abs/2406.19950}{{\ttfamily 2406.19950}}.

\end{thebibliography}\endgroup





\end{document}